%% file: main.tex
\documentclass[conference,10pt]{IEEEtran}
\usepackage{blindtext, graphicx, xcolor}

\usepackage[ruled,vlined,linesnumbered,noresetcount]{algorithm2e}
\usepackage{algpseudocode}

\usepackage[linguistics]{forest}

\newcommand{\minus}{\scalebox{0.75}[1.0]{$-$}}

\newcommand{\eat}[1]{}
\definecolor{purple}{rgb}{1,0,1}

\usepackage{hyperref}		\hypersetup{ colorlinks = true, urlcolor = blue, linkcolor = blue, citecolor = blue }

\definecolor{vt}{HTML}{892242}

\SetCommentSty{mycommfont}

\ifCLASSINFOpdf
\else
\fi

\begin{document}

\title{Full-speed Fuzzing: Reducing Fuzzing Overhead through Coverage-guided Tracing}

\author{Stefan Nagy \\ Virginia Tech \\ {snagy2@vt.edu} 
   \and Matthew Hicks \\ Virginia Tech \\ {mdhicks2@vt.edu} }

\maketitle

\makeatletter{}\begin{abstract}

Coverage-guided fuzzing is one of the most successful approaches for discovering software bugs and security vulnerabilities. 
Of its three main components: (1) test case generation, (2) code coverage tracing, and (3) crash triage, code coverage tracing is a dominant source of overhead. 
Coverage-guided fuzzers trace every test case's code coverage through either static or dynamic binary instrumentation, or more recently, using hardware support. 
Unfortunately, tracing \emph{all} test cases incurs significant performance penalties---even when the overwhelming majority of test cases and their coverage information are discarded because they do \emph{not} increase code coverage. 

To eliminate needless tracing by coverage-guided fuzzers, we introduce the notion of coverage-guided tracing. 
Coverage-guided tracing leverages two observations: (1) only a fraction of generated test cases increase coverage, and thus require tracing; and (2) coverage-increasing test cases become less frequent over time.  
Coverage-guided tracing encodes the current frontier of coverage in the target binary so that it self-reports when a test case produces new coverage---without tracing. 
This acts as a filter for tracing; restricting the expense of tracing to only coverage-increasing test cases. 
Thus, coverage-guided tracing trades increased time handling coverage-increasing test cases for decreased time handling non-coverage-increasing test cases.

To show the potential of coverage-guided tracing, we create an implementation based on the static binary instrumentor Dyninst called UnTracer.
We evaluate UnTracer using eight real-world binaries commonly used by the fuzzing community. 
Experiments show that after only an hour of fuzzing, UnTracer's average overhead is below 1\%, and after 24-hours of fuzzing, UnTracer approaches 0\% overhead, while tracing every test case with popular white- and black-box-binary tracers AFL-Clang, AFL-QEMU, and AFL-Dyninst incurs overheads of 36\%, 612\%, and 518\%, respectively. 
We further integrate UnTracer with the state-of-the-art hybrid fuzzer QSYM and show that in 24-hours of fuzzing, QSYM-UnTracer executes 79\% and 616\% more test cases than QSYM-Clang and QSYM-QEMU, respectively.

\end{abstract} 


\IEEEpeerreviewmaketitle

\makeatletter{}\section{Introduction}

Software vulnerabilities remain one of the most significant threats facing computer and information security \cite{noauthor_cve_2018}.
Real-world usage of weaponized software exploits by nation-states and independent hackers continues to expose the susceptibility of society's infrastructure to devastating cyber attacks.
For defenders, existing memory corruption and control-flow safeguards offer incomplete protection.
For software developers, manual code analysis does not scale to large programs.
Fuzzing, an automated software testing technique, is a popular approach for discovering software vulnerabilities due to its speed, simplicity, and effectiveness \cite{bounimova_billions_2012, serebryany_oss-fuzz_2017,swiecki_honggfuzz_2018, zalewski_american_2017}. 

At a high level, fuzzing consists of (1) generating test cases, (2) monitoring their effect on the target binary's execution, and (3) triaging bug-exposing and crash-producing test cases. 
State-of-the-art fuzzing efforts center on coverage-guided fuzzing \cite{zalewski_american_2017, swiecki_honggfuzz_2018, serebryany_continuous_2016, rawat_vuzzer:_2017, bohme_coverage-based_2016, li_steelix:_2017}, which augments execution with control-flow tracking apparatuses to trace test cases' code coverage (the exact code regions they execute).
Tracing enables coverage-guided fuzzers to focus mutation on a small set of unique test cases (those that reach previously-unseen code regions).
The goal being complete exploration of the target binary's code.

Code coverage is an abstract term that takes on three concrete forms in fuzzing literature: basic blocks, basic block edges, and basic block paths.
For white-box (source-available) binaries, code coverage is measured through instrumentation inserted at compile-time~\cite{swiecki_honggfuzz_2018, zalewski_american_2017, serebryany_continuous_2016}.
For black-box (source-unavailable) binaries, it is generally measured through instrumentation inserted dynamically~\cite{zalewski_american_2017, rawat_vuzzer:_2017} or statically through binary rewriting~\cite{talos-vulndev_afl-dyninst_2018}, or through instrumentation-free hardware-assisted tracing~\cite{schumilo_kafl:_2017, zhang_ptfuzz:_2018,swiecki_honggfuzz_2018}.

Tracing code coverage is costly---the largest source of time spent for most fuzzers---and the resulting coverage information is commonly discarded, as most test cases do \emph{not} increase coverage.
As our results in Section~\ref{sec:evaluation_tracing} show,  AFL~\cite{zalewski_american_2017}---one of the most popular fuzzers---faces tracing overheads as high as 1300\% for black-box binaries and as high as 70\% for white-box binaries.
These overheads are significant because, as experiments in Section~\ref{sec:perf_results} show, over 90\% of the time spent fuzzing involves executing and tracing test cases.
The problem with spending all this effort on coverage tracing is that most test cases and their coverage information are discarded; because, for most benchmarks in our evaluation, \textbf{less than 1 in 10,000} of all test cases are coverage-increasing.
Thus, the current practice of blindly tracing the coverage of every test case is incredibly wasteful.

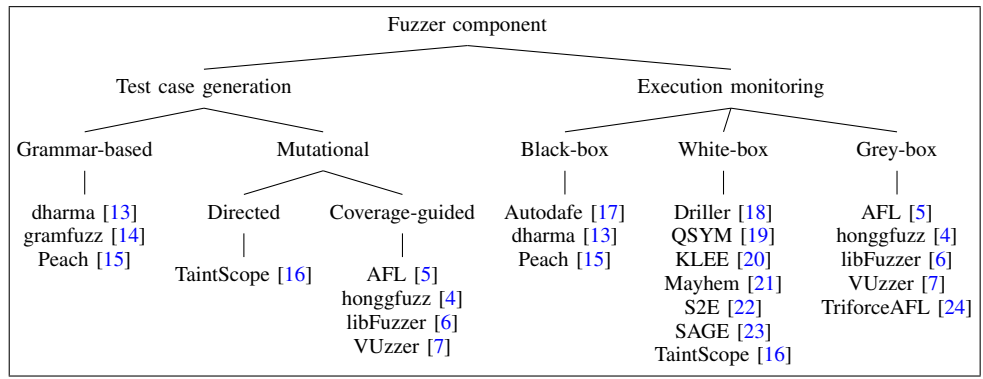
\begin{figure*}[!h]
        \footnotesize
    \centering
    \frame{\begin{forest}
    [Fuzzer component   
        [Test case generation
            [Grammar-based \\ [dharma \cite{mozilla_security_dharma:_2018}\\gramfuzz \cite{johnson_gramfuzz_2018}\\Peach \cite{eddington_peach_2018}]]
            [Mutational
                [Directed \\ [TaintScope \cite{wang_taintscope:_2010}]]
                [Coverage-guided \\ [AFL \cite{zalewski_american_2017}\\ honggfuzz \cite{swiecki_honggfuzz_2018}\\ libFuzzer \cite{serebryany_continuous_2016}\\ VUzzer \cite{rawat_vuzzer:_2017}]]
            ]
        ]
        [Execution monitoring
            [Black-box \\ [Autodafe \cite{vuagnoux_autodafe_2006}\\ dharma \cite{mozilla_security_dharma:_2018} \\ Peach \cite{eddington_peach_2018}]]
            [White-box \\ [Driller \cite{stephens_driller:_2016}\\QSYM \cite{yun_qsym:_2018}\\KLEE \cite{cadar_klee:_2008}\\Mayhem \cite{cha_unleashing_2012}\\S2E \cite{chipounov_s2e:_2011}\\SAGE \cite{godefroid_sage:_2012}\\TaintScope \cite{wang_taintscope:_2010}]]
            [Grey-box \\ [AFL \cite{zalewski_american_2017}\\honggfuzz \cite{swiecki_honggfuzz_2018}\\libFuzzer \cite{serebryany_continuous_2016}\\VUzzer \cite{rawat_vuzzer:_2017}\\TriforceAFL \cite{hertz_projecttriforce:_2017}]]
        ]
    ]
    \end{forest}}
    \caption{A taxonomy of popular fuzzers by test case generation and program analysis approaches.}
    \label{fig:fuzztaxonomy}
    \hrulefill
\end{figure*}

This paper introduces the idea of coverage-guided tracing, and its associated implementation UnTracer, targeted at reducing the overheads of coverage-guided fuzzers.
Coverage-guided tracing's goal is to restrict tracing to test cases \emph{guaranteed} to increase code coverage.
It accomplishes this by transforming the target binary so that it self-reports when a test case increases coverage.
We call such modified binaries \emph{interest oracles}. 
Interest oracles execute at native speeds because they eliminate the need for coverage tracing.
In the event that the interest oracle reports a test case is coverage-increasing, the test case is marked as coverage-increasing and conventional tracing is used to collect code coverage.
Portions of the interest oracle are then unmodified to reflect the additional coverage and the fuzzing process continues.
By doing this, coverage-guided tracing pays a high cost for handling coverage-increasing test cases (about 2x the cost of tracing alone in our experiments), for the ability to run all test cases (initially) at native speed.
To validate coverage-guided tracing and explore its tradeoffs on real-world software, we implement UnTracer.
UnTracer leverages the black-box binary instrumentor Dyninst~\cite{noauthor_dyninst_2018} to construct the interest oracle and tracing infrastructure.

We evaluate UnTracer alongside several coverage tracers used with the popular fuzzer AFL~\cite{zalewski_american_2017}. 
For tracing black-box binaries, we compare against the dynamic binary rewriter AFL-QEMU~\cite{zalewski_american_2017}, and the static binary rewriter AFL-Dyninst~\cite{noauthor_dyninst_2018}. 
For tracing white-box binaries, we compare against AFL-Clang~\cite{zalewski_american_2017}. 
To capture a variety of target binary and tracer behaviors, we employ a set of eight real-world programs of varying class and complexity (e.g., cryptography and image processing) that are common to the fuzzing community. 
In keeping with previous work, we perform evaluations for a 24-hour period and use 5 test case datasets per benchmark to expose the effects of randomness. 
Our results show UnTracer outperforms blindly tracing all test cases: UnTracer has an average run time overhead of 0.3\% across all benchmarks, while AFL-QEMU averages 612\% overhead, AFL-Dyninst averages 518\% overhead, and AFL-Clang averages 36\% overhead. 
Experimental results also show that the rate of coverage-increasing test cases rapidly approaches zero over time and would need to increase four orders-of-magnitude to ameliorate the need for UnTracer---even in a white-box tracing scenarios.
We further integrate UnTracer with the state-of-the-art hybrid fuzzer QSYM~\cite{yun_qsym:_2018}. 
Results show that QSYM-UnTracer averages 79\% and 616\% more executed test cases than QSYM-Clang and QSYM-QEMU, respectively.

In summary, this paper makes the following contributions:
\begin{itemize}
\item We introduce coverage-guided tracing: an approach for reducing fuzzing overhead by restricting tracing to coverage-increasing test cases.
\item We quantify the infrequency of coverage-increasing test cases across eight real-world applications.
\item We show that, for two coverage-guided fuzzers of different type: AFL (``blind'' test case generation) and Driller (``smart'' test case generation), they spend a majority of their time on tracing test cases.
\item We implement and evaluate UnTracer; UnTracer is our coverage-guided tracer based on the Dyninst black-box binary instrumentor. 
We evaluate UnTracer against three popular, state-of-the-art white- and black-box binary fuzzing tracing approaches: AFL-Clang (white-box), AFL-QEMU (black-box, dynamic binary rewriting), and AFL-Dyninst (black-box, static binary rewriting), using eight real-world applications.
\item We integrate UnTracer with the state-of-the-art hybrid fuzzer QSYM, and show that QSYM-UnTracer outperforms QSYM-Clang and QSYM-QEMU. 
\item We open-source our evaluation benchmarks~\cite{nagy_forte-fuzzbench:_2019}, experimental infrastructure~\cite{nagy_afl-fid:_2019}, and an AFL-based implementation of UnTracer~\cite{nagy_untracer-afl:_2019}.
\end{itemize} 

\makeatletter{}\section{Background}

In this section, we first discuss fuzzers' defining characteristics, and how they relate to UnTracer.
Second, we provide a detailed overview of coverage-guided fuzzing and how current fuzzers measure code coverage.
Third, we discuss related work on the performance of coverage tracing for fuzzing.
We conclude with our guiding research questions and principles.

\subsection{An Overview of Fuzzing}

Fuzzing is one of the most efficient and effective techniques for discovering software bugs and vulnerabilities.
Its simplicity and scalability have led to its widespread adoption among both bug hunters 
\cite{zalewski_american_2017, swiecki_honggfuzz_2018} and the software industry \cite{bounimova_billions_2012, serebryany_oss-fuzz_2017}. 
Fundamentally, fuzzers operate by generating enormous amounts of test cases, monitoring their effect on target binary execution behavior, and identifying test cases responsible for bugs and crashes.
Fuzzers are often classified by the approaches they use for test case generation and execution monitoring (Figure~\ref{fig:fuzztaxonomy}).

Fuzzers generate test cases using one of two approaches: grammar-based~\cite{godefroid_grammar-based_2008, mozilla_security_dharma:_2018, johnson_gramfuzz_2018, eddington_peach_2018} or mutational~\cite{sutton_fuzzing:_2007, zalewski_american_2017, swiecki_honggfuzz_2018, rawat_vuzzer:_2017, serebryany_continuous_2016}. 
Grammar-based generation creates test cases constrained by some pre-defined input grammar for the target binary. 
Mutational generation creates test cases using other test cases; in the first iteration, by mutating some valid ``seed'' input accepted by the target binary; and in subsequent iterations, by mutating prior iterations' test cases. 
For large applications, input grammar complexity can be burdensome, and for proprietary applications, input grammars are seldom available. 
For these reasons, most popular fuzzers are mutational.
Thus, coverage-guided tracing focuses on mutational fuzzing.

Most mutational fuzzers leverage program analysis to strategize which test cases to mutate.  
Directed fuzzers~\cite{bohme_directed_2017, ganesh_taint-based_2009} aim to reach specific locations in the target binary; thus they prioritize mutating test cases that seem to make progress toward those locations.
Coverage-guided fuzzers~\cite{zalewski_american_2017, swiecki_honggfuzz_2018, rawat_vuzzer:_2017, serebryany_continuous_2016} aim to explore the entirety of the target binary's code; thus they favor mutating test cases that reach new code regions. 
As applications of directed fuzzing are generally niche, such as taint tracking~\cite{wang_taintscope:_2010} or patch testing \cite{bohme_directed_2017}, coverage-guided fuzzing's wider scope makes it more popular among the fuzzing community~\cite{zalewski_american_2017, serebryany_continuous_2016, swiecki_honggfuzz_2018, serebryany_oss-fuzz_2017}.
Coverage-guided tracing is designed to enhance coverage-guided fuzzers.

Fuzzers are further differentiated based on the degree of program analysis they employ.
Black-box fuzzers~\cite{vuagnoux_autodafe_2006, mozilla_security_dharma:_2018, eddington_peach_2018} only monitor input/output execution behavior (e.g., crashes).
White-box fuzzers~\cite{godefroid_automated_2008, godefroid_sage:_2012, cha_unleashing_2012, stephens_driller:_2016, wang_taintscope:_2010, cadar_klee:_2008, chipounov_s2e:_2011} use heavy-weight program analysis for fine-grained execution path monitoring and constraint solving. 
Grey-box fuzzers~\cite{zalewski_american_2017, swiecki_honggfuzz_2018, rawat_vuzzer:_2017, serebryany_continuous_2016, hertz_projecttriforce:_2017, bohme_directed_2017, bohme_coverage-based_2016} are a tradeoff between both---utilizing lightweight program analysis (e.g., code coverage tracing).
Coverage-guided grey-box fuzzers are widely used in practice today; examples include VUzzer~\cite{rawat_vuzzer:_2017}, Google's libFuzzer~\cite{serebryany_continuous_2016}, honggfuzz~\cite{swiecki_honggfuzz_2018}, and AFL~\cite{zalewski_american_2017}.
Our implementation of coverage-guided tracing (UnTracer) is built atop the coverage-guided grey-box fuzzer AFL \cite{zalewski_american_2017}.

\begin{figure}[!t]
    \centering
    \frame{\includegraphics[width=1.0\columnwidth]{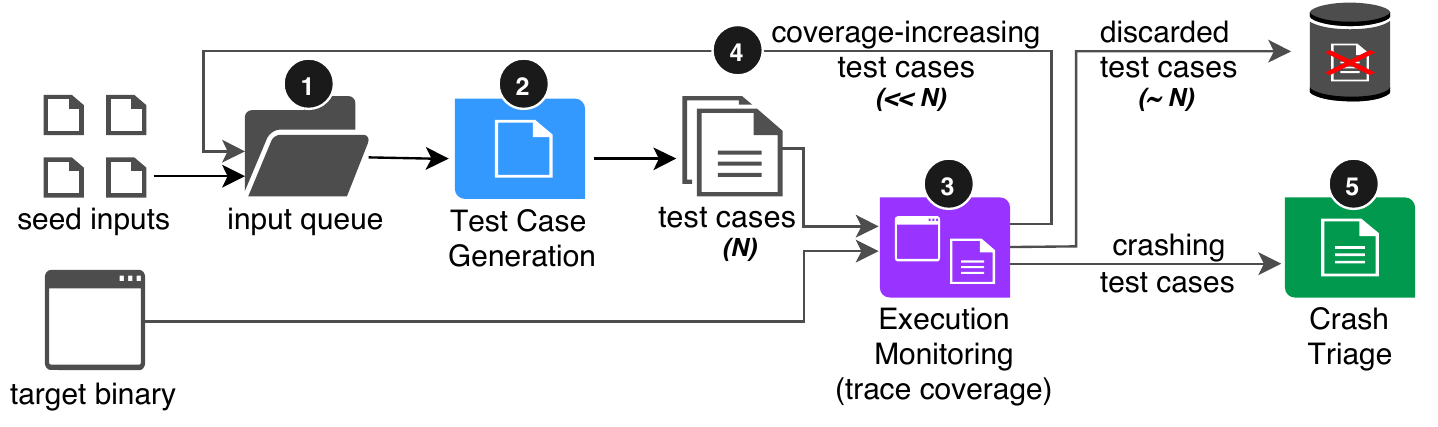}}
    \caption{High-level architecture of a coverage-guided mutational fuzzer.}
    \label{fig:cggbf}
    \hrulefill
\end{figure}

\subsection{Coverage-Guided Fuzzing}
\label{sec:background:cggbf}

Coverage guided fuzzing aims to explore the entirety of the target binary's code by maximizing generated test cases' code coverage.
Figure~\ref{fig:cggbf} highlights the high-level architecture of a coverage-guided mutational fuzzer. 
Given a target binary and some initial set of input seeds, $S$, fuzzing works as follows:

\begin{enumerate}
\item Queue all initial seeds\footnote{\emph{Seeds} refers to test cases used as the basis for mutation. In the first iteration, the seeds are generally several small inputs accepted by the target binary.} $s\in S$ for mutation.
\item{\textbf{test case generation:}} Select a queued seed and mutate it many times, producing test case set $T$.
\item{\textbf{Execution monitoring:}} For all test cases $t\in T$, trace their code coverage and look for crashes.
\item If a test case is \emph{coverage-increasing}, queue it as a seed, and prioritize it for the next round of mutation. Otherwise, discard it.
\item{\textbf{Crash triage:}} Report any crashing test cases.
\item Return to step 2 and repeat.
\end{enumerate}

\begin{figure}[!t]
    \centering
    \frame{\includegraphics[width=.5\columnwidth]{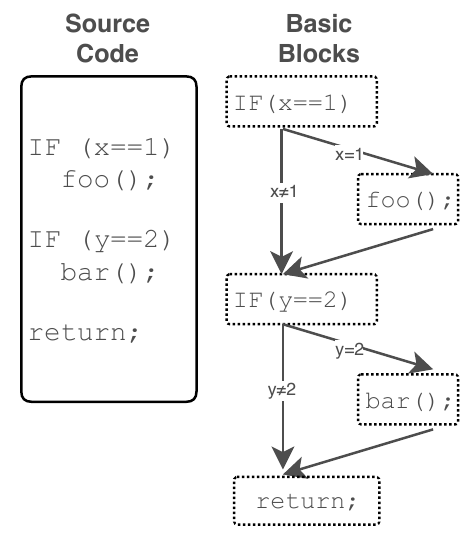}}
    \caption{An example of basic blocks in C code.}
    \label{fig:blocks}
    \hrulefill
\end{figure}

Coverage-guided fuzzers trace code coverage during execution via binary instrumentation \cite{zalewski_american_2017, serebryany_continuous_2016, swiecki_honggfuzz_2018}, system emulation \cite{zalewski_american_2017, schumilo_kafl:_2017, hertz_projecttriforce:_2017}, or hardware-assisted mechanisms \cite{schumilo_kafl:_2017, swiecki_honggfuzz_2018, zhang_ptfuzz:_2018}.
All coverage-guided fuzzers are based on one of three
metrics of code coverage: \emph{basic blocks}, basic block \emph{edges}, or basic block \emph{paths}. 
Basic blocks (Figure~\ref{fig:blocks}) refer to straight-lined sequences of code terminating in a control-flow transfer instruction (e.g., jumps or returns); they form the nodes of a program's control-flow graph. 

A basic block edge represents the actual control-flow transfer.
It is possible to represent edge coverage as a set of \texttt{(src , dest)} tuples, where \texttt{src} and \texttt{dest} are basic blocks.
Representing edge coverage this way (i.e., solely of basic blocks) allows edge coverage to be inferred from block coverage.
The caveat is that this requires prior elimination of all critical edges, i.e., edges whose starting/ending basic blocks have multiple outgoing/incoming edges, respectively (details in Section~\ref{sec:discuss_edge}).
honggFuzz~\cite{swiecki_honggfuzz_2018}, libFuzzer~\cite{serebryany_continuous_2016}, and AFL~\cite{zalewski_american_2017} are fuzzers that track coverage at edge granularity. 
honggFuzz and libFuzzer track edge coverage indirectly using block coverage, while AFL tracks edge coverage directly (although it stores the information approximately in a 64KB hash table~\cite{gan_collafl:_2018}).

To date, no fuzzers that we are aware of track coverage at path granularity, however, we can imagine future approaches leveraging Intel Processor Trace's~\cite{intel_intel_2017} ability to make tracking path coverage tractable. 
Thus, coverage-guided tracing complements coverage-guided fuzzers that trace block or edge coverage at block granularity.

\subsection{Coverage Tracing Performance}

Coverage-guided fuzzing of \emph{white-box (source-available) binaries} typically uses instrumentation inserted at compile/assembly-time~\cite{zalewski_american_2017, serebryany_continuous_2016, swiecki_honggfuzz_2018}, allowing for fast identification and modification of basic blocks from source.
AFL accomplishes this through custom GCC and Clang wrappers. 
honggfuzz and libFuzzer also provide their own Clang wrappers.
Fuzzing \emph{black-box (source-unavailable) binaries} is far more challenging, as having no access to source code requires costly reconstruction of binary control-flow.
VUzzer \cite{rawat_vuzzer:_2017} uses PIN \cite{luk_pin:_2005} to dynamically (during run-time) instrument black-box binaries. 
AFL's QEMU user-mode emulation also instruments dynamically, but as our experiments show (Section~\ref{sec:evaluation_tracing}), it incurs overheads as high as 1000\% compared to native execution.
To address the weakness of dynamic rewriting having to translate basic blocks in real-time---potentially multiple times---Cisco-Talos provides a static binary rewriter AFL-Dyninst~\cite{talos-vulndev_afl-dyninst_2018}. 
While previous work shows AFL-Dyninst significantly outperforms AFL-QEMU on select binaries~\cite{nikolic_guided_2016}, results in Section~\ref{sec:evaluation_tracing} suggest that the performance gap is much narrower.

\begin{table*}
\centering
\footnotesize
\begin{tabular}{ l c c c c c c c c c }
 & \texttt{bsdtar} & \texttt{cert-basic} & \texttt{cjson} & \texttt{djpeg} & \texttt{pdftohtml} & \texttt{readelf} & \texttt{sfconvert} & \texttt{tcpdump} & \textbf{avg.} \\ 
\hline
\hline
\textbf{AFL-Clang} & 89.4 & 91.9 & 86.0 & 94.7 & 98.4 & 86.9 & 99.2 & 88.3 & 91.8 \\
\textbf{AFL-QEMU} & 95.7 & 98.9 & 95.7 & 97.8 & 99.5 & 96.5 & 98.6 & 95.8 & 97.3 \\
\\
 & \texttt{CADET\_1} & \texttt{CADET\_3} & \texttt{CROMU\_1} & \texttt{CROMU\_2} & \texttt{CROMU\_3} & \texttt{CROMU\_4} & \texttt{CROMU\_5} & \texttt{CROMU\_6} & \textbf{avg.}\\ 
\hline
\hline
\textbf{Driller-AFL} & 97.6 & 97.1 & 96.0 & 94.9 & 96.0 & 93.1 & 97.5 & 94.9 & 95.9 \\
\end{tabular}
\vspace{0.2cm}
\caption{Per-benchmark percentages of total fuzzing runtime spent on test case execution and coverage tracing by AFL-Clang and AFL-QEMU (``blind'' fuzzing), and Driller-AFL (``smart'' fuzzing). We run each fuzzer for one hour per benchmark.
}
\label{tab:perfresults1}
\hrulefill{}
\end{table*}

\begin{table*}
\centering
\footnotesize
\begin{tabular}{ l  c c c c c c c c c }
 & \texttt{bsdtar} & \texttt{cert-basic} & \texttt{cjson} & \texttt{djpeg} & \texttt{pdftohtml} & \texttt{readelf} & \texttt{sfconvert} & \texttt{tcpdump} & \textbf{avg.}\\ 
\hline
\hline
\textbf{AFL-Clang} & 1.63E\minus5 & 4.47E\minus5 & 2.78E\minus6 & 4.30E\minus5 & 1.42E\minus4 & 7.43E\minus5 & 8.77E\minus5 & 8.55E\minus5 & 6.20E\minus5 \\
\textbf{AFL-QEMU} & 3.34E\minus5 & 4.20E\minus4 & 1.41E\minus5 & 1.09E\minus4 & 6.74E\minus4 & 2.28E\minus4 & 4.25E\minus4 & 1.55E\minus4 & 2.57E\minus4 \\
\\
 & \texttt{CADET\_1} & \texttt{CADET\_3} & \texttt{CROMU\_1} & \texttt{CROMU\_2} & \texttt{CROMU\_3} & \texttt{CROMU\_4} & \texttt{CROMU\_5} & \texttt{CROMU\_6} & \textbf{avg.}\\ 
\hline
\hline
\textbf{Driller-AFL} & 2.70E\minus5 & 4.00E\minus4 & 2.06E\minus5 & 2.67E\minus5 & 2.33E\minus5 & 8.65E\minus7 & 1.61E\minus5 & 8.45E\minus6 & 6.53E\minus5 \\
\end{tabular}
\vspace{0.2cm}
\caption{Per-benchmark rates of coverage-increasing test cases out of all test cases generated in one hour by AFL-Clang and AFL-QEMU (``blind'' fuzzing), and Driller-AFL (``smart'' fuzzing). 
}
\label{tab:perfresults2}
\hrulefill{}
\end{table*}

\subsection{Focus of this Paper}

A characteristic of coverage-guided fuzzing is the coverage tracing of \emph{all} generated test cases.
Though ``smarter'' fuzzing efforts generate coverage-increasing test cases with higher frequency, results in Section~\ref{sec:perf} show that only a small percentage of all test cases are coverage-increasing.
We draw inspiration from Amdahl's Law~\cite{gustafson_reevaluating_1988}, realizing that the common case---the tracing of non-coverage-increasing test cases---presents an opportunity to substantially improve the performance of coverage-guided fuzzing.
Thus we present coverage-guided tracing, which restricts tracing to only coverage-increasing test cases. 
Our implementation, UnTracer, is a coverage-guided tracing framework for coverage-guided fuzzers.

\makeatletter{}\section{Impact of Discarded test cases}
\label{sec:perf}

Traditional coverage-guided fuzzers (e.g., AFL~\cite{zalewski_american_2017}, libFuzzer~\cite{serebryany_continuous_2016}, and honggfuzz~\cite{swiecki_honggfuzz_2018}) rely on ``blind'' (random mutation-based) test case generation; coverage-increasing test cases are preserved and prioritized for future mutation, while the overwhelming majority are non-coverage-increasing and discarded along with their coverage information.
To reduce rates of non-coverage-increasing test cases, several white-box and grey-box fuzzers employ ``smart'' test case generation. Smart mutation leverages source analysis (e.g., symbolic execution~\cite{stephens_driller:_2016}, program state~\cite{li_steelix:_2017}, and taint tracking~\cite{chen_angora:_2018, rawat_vuzzer:_2017}) to generate a higher proportion of coverage-increasing test cases.
However, it is unclear if such fuzzers spend significantly more time on test case generation than on test case execution/coverage tracing or how effective smart mutation is at increasing the rate of coverage-increasing test cases.

In this section, we investigate the performance impact of executing/tracing non-coverage-increasing test cases in two popular state-of-the-art fuzzers---AFL (blind test case generation) \cite{zalewski_american_2017} and Driller (smart test case generation) \cite{stephens_driller:_2016}.
We measure the runtime spent by both AFL and Driller on executing/tracing test cases across eight binaries, for one hour each, and their corresponding rates of coverage-increasing test cases.
Below, we highlight the most relevant implementation details of both fuzzers regarding test case generation and coverage tracing, and our experimental setup.

\subsubsection*{AFL}

AFL \cite{zalewski_american_2017} is a ``blind'' fuzzer as it relies on random mutation to produce coverage-increasing (coverage-increasing) test cases, which are then used during mutation.\footnote{A second, less-relevant factor influencing AFL's test case mutation priority is test case size. For two test cases exhibiting identical code coverage, AFL will prioritize the test case with smaller filesize \cite{zalewski_american_2017}.}
AFL traces test case coverage using either QEMU-based dynamic instrumentation for black-box binaries or assembly/compile-time instrumentation for white-box binaries.
We cover both options by evaluating AFL-QEMU and AFL-Clang.

\subsubsection*{Driller}
Driller \cite{stephens_driller:_2016} achieves ``smart'' test case generation by augmenting blind mutation with selective concolic execution---solving path constraints symbolically (instead of by brute-force). 
Intuitively, Driller aims to outperform blind fuzzers by producing fewer non-coverage-increasing test cases; its concolic execution enables penetration of path constraints where blind fuzzers normally stall.
We evaluate Driller-AFL (aka ShellPhuzz \cite{shellphish_shellphuzz_2018}).
Like AFL, Driller-AFL also utilizes QEMU for black-box binary coverage tracing. 

\subsection{Experimental Setup}
For AFL-Clang and AFL-QEMU we use the eight benchmarks from our evaluation in Section~\ref{sec:evaluation_tracing}.
As Driller currently only supports benchmarks from the DARPA Cyber Grand Challenge (CGC)~\cite{noauthor_darpa_2018}, we evaluate Driller-AFL on eight pre-compiled~\cite{shoshitaishvili_cgc_2017} CGC binaries. 
We run all experiments on the same setup as our performance evaluation (Section~\ref{sec:evaluation_tracing}).

To measure each fuzzer's execution/tracing time, we insert timing code in AFL's test case execution function (\texttt{run\_target()}).
As timing occurs per-execution, this allows us to also log the total number of test cases generated.
We count each fuzzer's coverage-increasing test cases by examining its AFL queue directory and counting all saved test cases AFL appends with tag \texttt{+cov}---its indicator that the test case increases code coverage.

\subsection{Results}
\label{sec:perf_results}

As shown in Table~\ref{tab:perfresults1}, both AFL and Driller spend the majority of their runtimes on test case execution/coverage tracing across all benchmarks: AFL-Clang and AFL-QEMU average 91.8\% and 97.3\% of each hour, respectively, while Driller-AFL averages 95.9\% of each hour.
Table~\ref{tab:perfresults2} shows each fuzzer's rate of coverage-increasing test cases across all one-hour trials.
On average, AFL-Clang and AFL-QEMU have .0062\% and .0257\% coverage-increasing test cases out of all test cases generated in one hour, respectively. 
Driller-AFL averages .00653\% coverage-increasing test cases out of all test cases in each one hour trial. These results show that coverage-guided fuzzers AFL (blind) and Driller (smart)---despite adopting different test case generation methodologies---\emph{both spend the majority of their time executing and tracing the coverage of non-coverage-increasing test cases.}

\makeatletter{}\begin{figure*}[!t]
    \centering
    \frame{\includegraphics[width=0.75\textwidth]{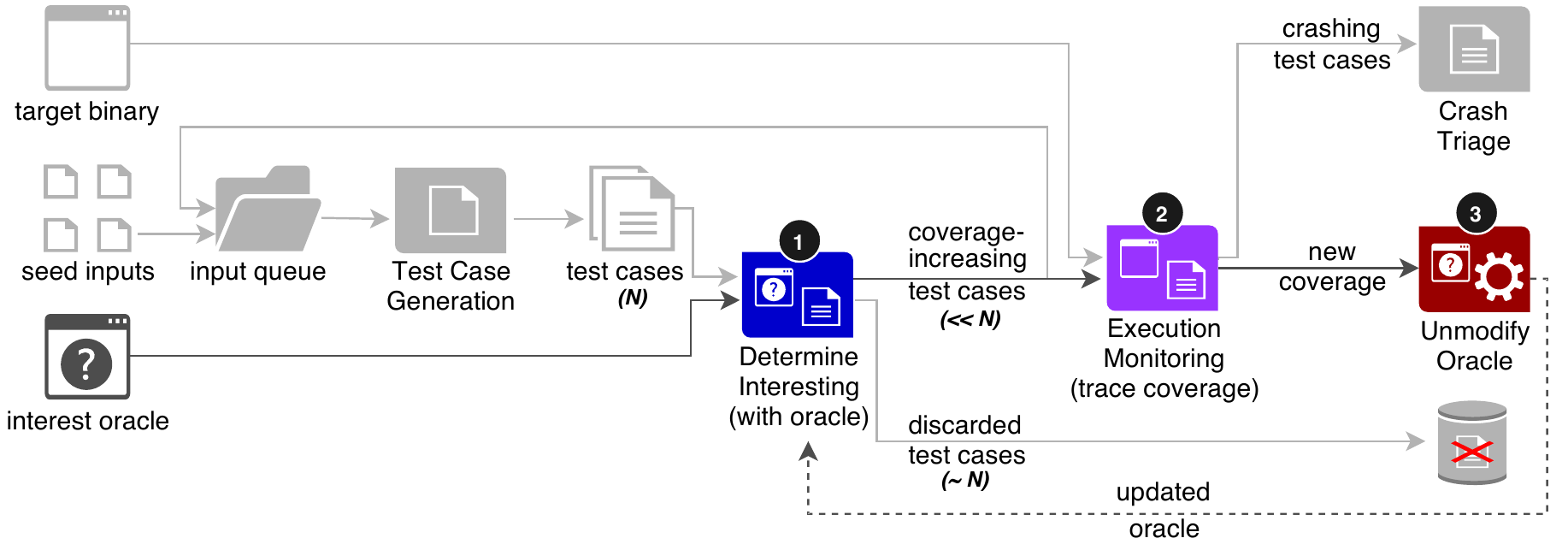}}
    \caption{Visualization of how coverage-guided tracing augments the workflow of a conventional coverage-guided grey-box fuzzer (e.g., AFL \cite{zalewski_american_2017}). Coverage-guided tracing can also be similarly adapted into coverage-guided white-box fuzzers (e.g., Driller \cite{stephens_driller:_2016}).}
    \label{fig:cgt-workflow}
    \hrulefill
\end{figure*}

\section{Coverage-guided Tracing}

Current coverage-guided fuzzers trace \emph{all} generated test cases to compare their individual code coverage to some accumulated \emph{global} coverage. 
Test cases with \emph{new} coverage are retained for mutation and test cases without new coverage are discarded along with their coverage information. 
In Section~\ref{sec:perf}, we show that two coverage-guided fuzzers of different type---AFL (``blind'') and Driller (``smart'')---both spend the majority of their time executing/tracing non-coverage-increasing test cases.
\emph{Coverage-guided tracing} aims to trace \emph{fewer} test cases by restricting tracing to \emph{only} coverage-increasing test cases.

\subsection{Overview}

Coverage-guided tracing introduces an intermediate step between test case generation and code coverage tracing: the \emph{interest oracle}.
An interest oracle is a modified version of the target binary, where a pre-selected software interrupt is inserted via overwriting at the start of each uncovered basic block.
Interest oracles restrict tracing to only coverage-increasing test cases as follows: test cases that trigger the oracle's interrupt are marked coverage-increasing, and then traced.
As new basic blocks are recorded, their corresponding interrupts are removed from the oracle binary (\emph{unmodifying})---making it increasingly mirror the original target.
As this process repeats, only test cases exercising \emph{new} coverage trigger the interrupt---thus signaling them as coverage-increasing.

As shown in Figure~\ref{fig:cgt-workflow}, coverage-guided tracing augments conventional coverage-guided fuzzing by doing the following:

\begin{enumerate}
    \item \textbf{Determine Interesting:} Execute a generated test case against the \emph{interest oracle}. If the test case triggers the interrupt, mark it as coverage-increasing. Otherwise, return to step 1.
    \item \textbf{Full Tracing:} For every coverage-increasing test case, trace its full code coverage. 
    \item \textbf{Unmodify Oracle:} For every \emph{newly-visited} basic block in the test case's coverage, remove its corresponding interrupt from the interest oracle.
    \item Return to step 1.
\end{enumerate}

\subsection{The Interest Oracle}
In coverage-guided tracing, interest oracles sit between test case generation and coverage tracing---acting as a mechanism for filtering-out non-coverage-increasing test cases from being traced.
Given a target binary, an interest oracle represents a modified binary copy with a software interrupt signal overwriting the start of each basic block.
A test case is marked \emph{coverage-increasing} if it triggers the interrupt---meaning it has entered some previously-uncovered basic block.
Coverage-increasing test cases are then traced for their \emph{full} coverage, and their newly-covered basic blocks are \emph{unmodified} (interrupt removed) in the interest oracle.

Interest oracle construction requires prior identification of the target binary's basic block addresses.
Several approaches for this exist in literature~\cite{kinder_abstract_2009, theiling_extracting_2000, kastner_generic_2002}, and tools like angr~\cite{shoshitaishvili_sok:_2016} and Dyninst~\cite{noauthor_dyninst_2018} can also accomplish this via static analysis.
Inserting interrupts is trivial, but bears two caveats: first, while any interrupt signal can be used, it should avoid conflicts with other signals central to fuzzing (e.g., those related to crashes or bugs); second, interrupt instruction size must not exceed any candidate basic block's size (e.g., one-byte blocks cannot accommodate two-byte interrupts).

\subsection{Tracing}
Coverage-guided tracing derives coverage-increasing test cases' \emph{full} coverage through a separate, tracing-only version of the target.
As interest oracles rely on block-level binary modifications, code coverage tracing must also operate at block-level.
Currently, block-level tracing can support either block coverage~\cite{rawat_vuzzer:_2017}, or---if all critical edges are mitigated---edge coverage~\cite{swiecki_honggfuzz_2018, serebryany_continuous_2016}.
Thus, coverage-guided tracing is compatible with most existing tracing approaches.

\subsection{Unmodifying}
Coverage-guided tracing's unmodify routine removes oracle interrupts in newly-covered basic blocks.
Given a target binary, an interest oracle, and a list of newly-covered basic blocks, unmodifying overwrites each block's interrupt with the instructions from the original target binary.

\subsection{Theoretical Performance Impact}
\begin{figure}
    \centering
    \frame{\includegraphics[width=.8\columnwidth]{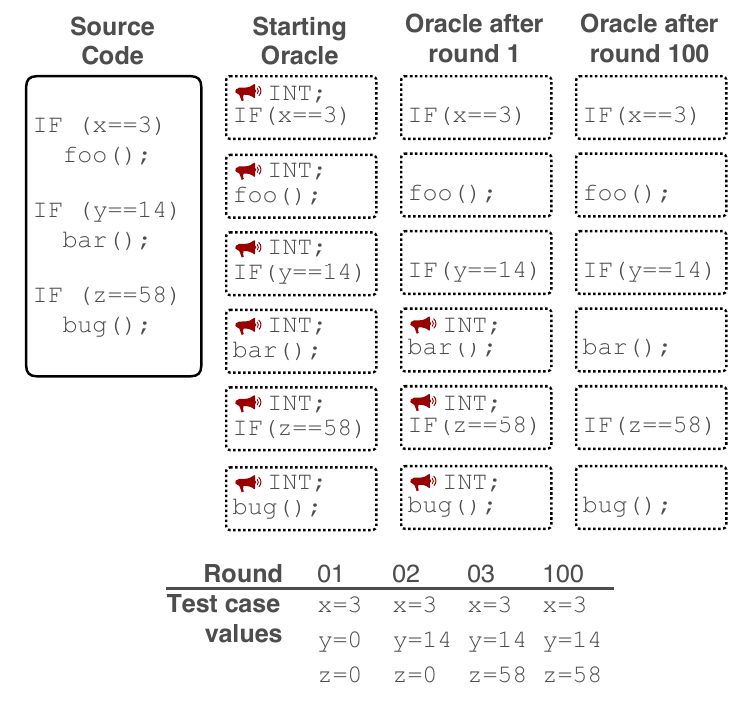}}
    \caption{An example of the expected evolution of a coverage-guided tracing interest oracle's basic blocks alongside its original source code. Here, \texttt{INT} denotes an oracle interrupt.
    For simplicity, this diagram depicts interrupts as inserted; however, in coverage-guided tracing, the interrupts instead overwrite the start of each basic block. Unmodifying basic blocks consists of resetting their interrupt-overwritten byte(s) to their original values.
                }
    \label{fig:cgt-blocks}
    \hrulefill
\end{figure}

Over time, a growing number of coverage-increasing test cases causes more of the oracle's basic blocks to be unmodified (Figure~\ref{fig:cgt-blocks})---thus reducing the dissimilarity between oracle and target binaries.
As the oracle more closely resembles the target, it becomes less likely that a test case will be coverage-increasing (and subsequently traced). Given that non-coverage-increasing test cases execute at the same speed for both the original and the oracle binaries, as fuzzing continues, coverage-guided tracing's overall performance approaches 0\% overhead.

\makeatletter{}\section{Implementation: UnTracer}
\label{sec:implementation}

\begin{figure*}
    \centering
    \frame{\includegraphics[width=0.7\textwidth]{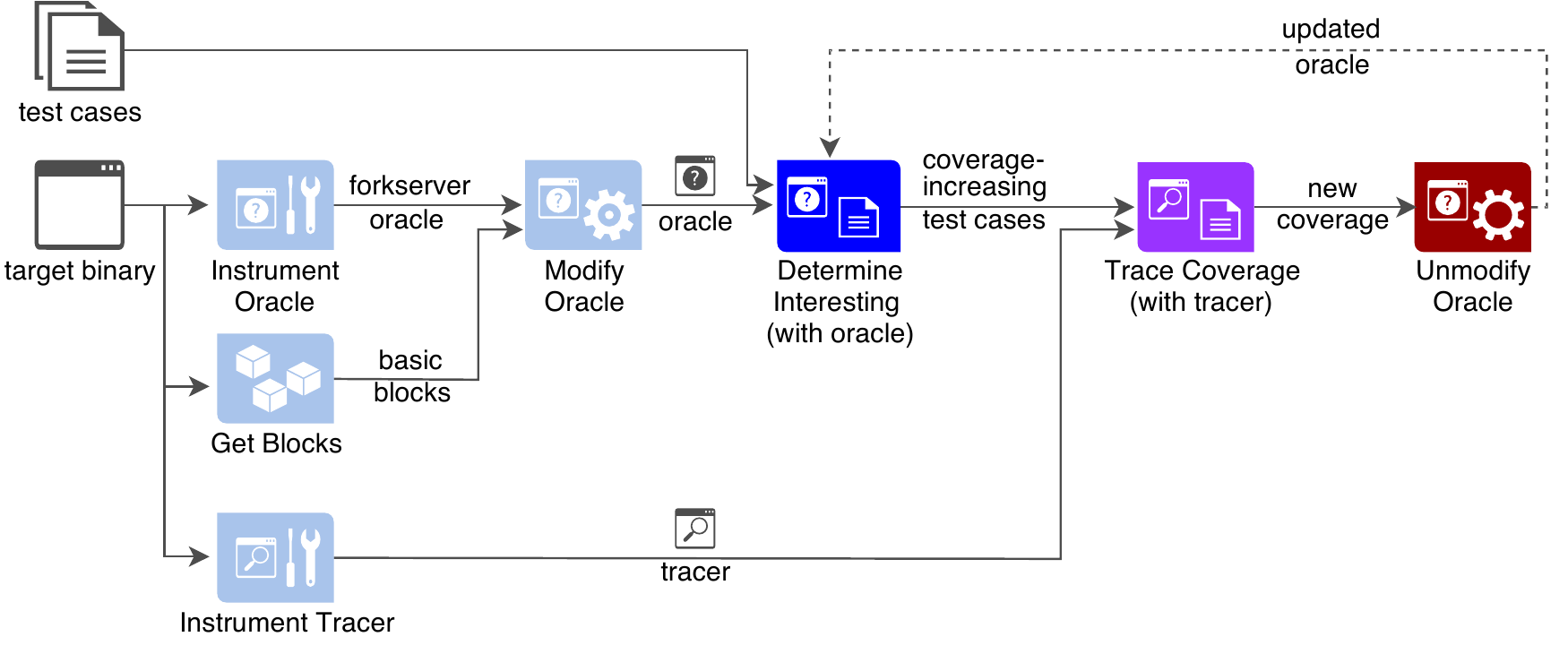}}
    \caption{UnTracer's workflow. Not shown is test case generation, or starting/stopping forkservers.}
        \label{fig:untracer-workflow}
    \hrulefill
\end{figure*}

Here we introduce \emph{UnTracer}, our implementation of coverage-guided tracing.
Below, we offer an overview of UnTracer's algorithm and discuss its core components in detail.

\subsection{UnTracer Overview}
UnTracer is built atop a modified version of the coverage-guided grey-box fuzzer, AFL 2.52b~\cite{zalewski_american_2017}, which we selected due to both its popularity in the fuzzing literature~\cite{peng_t-fuzz:_2018, stephens_driller:_2016, lemieux_perffuzz:_2018, bohme_coverage-based_2016, bohme_directed_2017, hertz_projecttriforce:_2017, li_steelix:_2017, wang_skyfire:_2017} and its open-source availability.
Our implementation consists of 1200 lines of C and C++ code.
UnTracer instruments two separate versions of the target binary---an \emph{interest oracle} for identifying coverage-increasing test cases, and a \emph{tracer} for identifying new coverage.
As AFL utilizes a forkserver execution model~\cite{zalewski_fuzzing_2014}, we incorporate this in both UnTracer's oracle and tracer binaries.

Algorithm~\ref{alg:untracer} shows the steps UnTracer takes, as integrated with AFL.
After AFL completes its initial setup routines (e.g., creating working directories and file descriptors) (line 1), UnTracer instruments both the oracle and tracer binaries (lines 2--3); the oracle binary gets a forkserver while the tracer binary gets a forkserver and basic block-level instrumentation for coverage tracing.
As the oracle relies on block-level software interrupts for identifying coverage-increasing test cases, UnTracer first identifies all basic blocks using static analysis (line 5); then, UnTracer inserts the interrupt at the start of every basic block in the oracle binary (lines 6--8). 
To initialize both the oracle and tracer binaries for fuzzing, UnTracer starts their respective forkservers (lines 9--10).
During AFL's main fuzzing loop (lines 11--23), UnTracer executes every AFL-generated test case (line 12) on the oracle binary (line 13).
If any test case triggers an interrupt, UnTracer marks it as coverage-increasing (line 14) and uses the tracer binary to collect its coverage (line 15).
We then stop the forkserver (line 16) to unmodify every newly-covered basic block (lines 17-19)---removing their corresponding oracle interrupts; this ensures only future test cases with new coverage will be correctly identified as coverage-increasing.
After all newly-covered blocks have been unmodified, we restart the updated oracle's forkserver (line 20).
Finally, AFL completes its coverage-increasing test case handling routines (e.g., queueing and prioritizing for mutation) (line 21) and fuzzing moves onto the next test case (line 12). Figure~\ref{fig:untracer-workflow} depicts UnTracer's architecture.

\renewcommand{\algorithmicrequire}{\textbf{Data:}}

\begin{algorithm}[!t]
    \footnotesize
    \DontPrintSemicolon
    \SetNoFillComment
    \SetAlgoLined
    \KwIn{$P$: the target program}
    \KwData{$b$: a basic block\newline$B$: a set of basic blocks\newline$i$: an AFL-generated test case\newline $\Phi$: the set of all coverage-increasing test cases}
    \enspace
    \textsc{AFL\_Setup()}\;
    \tcp{Instrument oracle and tracer binaries}
    $P_O \leftarrow \textsc{instOracle}(P)$\; 
    $P_T \leftarrow \textsc{instTracer}(P)$\;

    \tcp{Find and modify all of oracle's blocks}
    $B = \emptyset$\;
    $B \leftarrow \textsc{getBasicBlocks}(P)$\;
    \For{$b \in B$}{
        $\textsc{modifyOracle}(b)$\;
    }
    \tcp{Start oracle and tracer forkservers}
    $\textsc{startForkserver}(P_O)$\;
    $\textsc{startForkserver}(P_T)$\;
   
    \enspace
    \tcp{Main fuzzing loop}
    \While{1}{
        $i \leftarrow \textsc{AFL\_WriteTotest case()}$\;
        
        \If{$P_O(i) \rightarrow \texttt{INTERRUPT}$}{
            \tcp{The test case is \emph{coverage-increasing}}
            $\Phi.\textsc{add}(i)$\;

            \tcp{Trace test case's new coverage}
            $B_{trace} \leftarrow  \textsc{getTrace}(P_T(i))$\;
            
            \tcp{Kill oracle before unmodifying} $\textsc{stopForkserver}(P_O)$\;            
            
            \tcp{Unmodify test case's new coverage}
            \For{$b \in B_{trace}$}{
                $\textsc{unmodifyOracle}(b)$\;
            }
            \tcp{Restart oracle before continuing} $\textsc{startForkserver}(P_O)$\;            
            \textsc{AFL\_HandleCoverageIncreasing()}\;
        }
    }
    \caption{The UnTracer algorithm integrated in AFL.}
    \label{alg:untracer}
\end{algorithm}

\subsection{Forkserver Instrumentation}
\label{sec:implementation:forkserver}
During fuzzing, both UnTracer's oracle and tracer binaries are executed many times; the oracle executes all test cases to determine which are coverage-increasing and the tracer executes all coverage-increasing test cases to identify new coverage.
In implementing UnTracer, we aim to optimize execution speeds of both binaries.
Like other AFL tracers, UnTracer incorporates a \emph{forkserver} execution model \cite{zalewski_fuzzing_2014} in its tracer binary, as well as in its oracle binary.
By launching new processes via \texttt{fork()}, the forkserver avoids repetitive process initialization---achieving significantly faster speeds than traditional \texttt{execve()}-based execution.
Typically, instrumentation first inserts a forkserver function in a binary's \texttt{.text} region, and then links to it a callback in the first basic block of function \texttt{<main>}.
In the tracer binary, we already use Dyninst's static binary rewriting for black-box binary instrumentation, so we use that same technique for the forkserver.

For the oracle binary, our initial approach was to instrument it using Dyninst. Unfortunately, preliminary evaluations revealed several performance problems.\footnote{\label{note:dyninst}We made Dyninst developers aware of several performance issues---specifically, excessive function calls (e.g., to \texttt{\_\_dl\_relocate\_object}) after exiting the forkserver function. While they confirmed that this behavior is unexpected, they were unable to remedy these issues before publication.}
Since the oracle executes every test case, it is performance critical. 
To avoid Dyninst's limitations, we leverage AFL's assembly-time instrumentation to insert the forkserver in the oracle binary, since it closely mimics the outcome of black-box binary rewriters.

\subsection{Interest Oracle Binary}
The oracle is a modified version of the target binary that adds the ability to self-report coverage-increasing test cases through the insertion of software interrupts at the start of each \emph{uncovered} basic block.
Thus, if a test case triggers the interrupt, it has exercised some \emph{new} basic block and is marked as coverage-increasing.
Oracle binary construction requires prior knowledge of the target binary's basic block addresses.
We leverage Dyninst's static control-flow analysis to output a list of basic blocks, then iterate through that list in using binary file IO to insert the interrupts.
To prevent interrupts triggering before forkserver initialization, we do not consider functions executed prior to the forkserver callback in \texttt{<main>} (e.g., \texttt{<\_start>, <\_libc\_start\_main>, <\_init>, <frame\_dummy>}).

We use \texttt{SIGTRAP} for our interrupt for two reasons: (1) it has long been used for fine-grain execution control and analysis (e.g., \texttt{gdb}~\cite{stallman_debugging_1988, brown_architecture_2012} and kernel-probing~\cite{keniston_kernel_2016, hiramatsu_djprobe--kernel_2007}); and (2) its binary representation---\texttt{0xCC}---is one byte long, making it possible to overwrite basic blocks of all sizes.

\subsection{Tracer Binary}
If the oracle determines a test case to be coverage-increasing, UnTracer extracts its new code coverage by executing it on a separate \emph{tracer} binary---a coverage tracing-instrumented version of the target binary.
We utilize Dyninst to statically instrument the tracer with a forkserver for fast execution, and coverage callbacks inserted in each of its basic blocks for coverage tracing.
Upon execution, a basic block's callback appends its corresponding basic block address to a trace file located in UnTracer's working directory.

In an early version of UnTracer, we observed that coverage traces containing repeatedly-executing basic blocks add significant overhead in two ways: first, recording individual blocks multiple times---a common occurrence for binaries exhibiting looping behavior---slowed UnTracer's trace \emph{writing} operations; second, trace \emph{reading} is also penalized, as subsequent block-level unmodification operations are forced to process any repeatedly-executing basic blocks.
To mitigate such overhead, we optimize tracing to only record \emph{uniquely}-covered basic blocks as follows:
in the tracer forkserver, we initialize a global hashmap data structure to track all covered basic blocks unique to each trace; as each tracing child is forked, it inherits the initial hashmap; upon a basic block's execution, its callback utilizes hashmap lookup to determine if the block has been previously covered in the current execution; if not, the callback updates the current trace log and updates the hashmap.
With this optimization, for each coverage-increasing test case, UnTracer records a set of all uniquely-covered basic blocks, thus reducing the overhead resulting from logging, reading, and processing the same basic block multiple times.

\subsection{Unmodifying the Oracle}
When a test case triggers the oracle's software interrupt, it is marked as coverage-increasing and UnTracer removes its interrupts from its newly-covered basic blocks to ensure no future test case with the non-new coverage is marked coverage-increasing.
For each newly-covered basic block reported in an coverage-increasing test case's trace log, UnTracer replaces the inserted interrupt with the original byte found in the target binary---effectively resetting it to its pre-modified state. 
Doing so means any future test cases executing this basic block will no longer trigger the interrupt and subsequently not be misidentified as coverage-increasing.

We observe that even coverage-increasing test cases often have significant overlaps in coverage.
This causes UnTracer to attempt unmodifying many already-unmodified basic blocks, resulting in high overhead.
To mitigate this, we introduce a hashmap data structure for tracking \emph{global} coverage.
Much like the hashmap used for per-trace redundant basic block filtering, before unmodifying any basic block from the trace log, UnTracer determines if the block has been seen in any previous trace via hashmap lookup.
If so, the basic block is skipped.
If not, its interrupt is removed, and the basic block is added to the hashmap.
Thus, global coverage tracking ensures that only newly-covered basic blocks are processed. 

\makeatletter{}\makeatletter{}\begin{table*}[!t]

\centering
\small\begin{tabular}{ l l r c r r r r}
\textbf{Package} & \textbf{Benchmark} & \textbf{Version} & \textbf{Class} & \textbf{Basic Blocks} &  \textbf{Test Cases ($\cdot10^6$)} & \multicolumn{1}{p{1.8cm}}{\raggedleft \textbf{Coverage-increasing Ratio}} & \multicolumn{1}{p{1.3cm}}{\raggedleft \textbf{500ms Timeouts}} \\
\hline
\hline
\texttt{libarchive} & \texttt{bsdtar} & 3.3.2 & archiv & 31379 & 21.06 & 1.47E\minus5 & 0\\

\texttt{libksba} & \texttt{cert-basic} & 1.3.5 & crypto & 9958 & 10.73 & 1.50E\minus5 & 0\\

\texttt{cjson} & \texttt{cjson} & 1.7.7 & web & 1447 & 25.62 & 1.48E\minus5 & 0\\

\texttt{libjpeg} & \texttt{djpeg} & 9c & image & 4844 & 14.53 & 1.33E\minus5 & 12133\\\hline

\texttt{poppler} & \texttt{pdftohtml} & 0.22.5 & doc & 54596 & 1.21 & 7.85E\minus5 & 0\\

\texttt{binutils} & \texttt{readelf} & 2.30 & dev & 21249 & 14.89 & 8.98E\minus5 & 0\\

\texttt{audiofile} & \texttt{sfconvert} & 0.2.7 & audio & 5603 & 10.17 & 3.91E\minus2 & 1137609\\

\texttt{tcpdump} & \texttt{tcpdump} & 4.9.2 & net & 33743 & 27.14 & 3.73E\minus5 & 0 \\

\end{tabular}

\vspace{0.2cm}
\caption{Information on the eight benchmarks used in our evaluation in Sections~\ref{sec:evaluation_tracing} and ~\ref{sec:evaluation_fuzzing} and averages over 5 24-hour datasets for each benchmark.}
\label{tab:binaries}
\hrulefill{}
\end{table*} 

\section{Tracing-only Evaluation}
\label{sec:evaluation_tracing}

Our evaluation compares UnTracer against tracing all test cases with three widely used white- and black-box binary fuzzing tracing approaches---AFL-Clang (white-box)~\cite{zalewski_american_2017}, AFL-QEMU (black-box dynamically-instrumented)~\cite{zalewski_american_2017}, and AFL-Dyninst (black-box statically-instrumented)~\cite{talos-vulndev_afl-dyninst_2018} on eight real-world benchmarks of different type. 

Our experiments answer the following questions:
\begin{enumerate}
    \item How does UnTracer (\emph{coverage-guided tracing}) perform compared to tracing all test cases?
    \item What factors contribute to UnTracer's overhead?
    \item How is UnTracer's overhead impacted by the rate of coverage-increasing test cases?
\end{enumerate}

\subsection{Evaluation Overview}

We compare UnTracer's performance versus popular white- and black-box fuzzing tracing approaches: AFL-Clang, AFL-QEMU, and AFL-Dyninst.
These tracers all work with the same fuzzer, AFL, and they cover the tracing design space including working with white- and black-box binaries as well as static and dynamic binary rewriting.
Our evaluations examine each tracer's overhead on eight real-world, open-source benchmarks of different type, common to the fuzzing community.
Table~\ref{tab:binaries} provides benchmark details.
To smooth the effects of randomness and ensure the most fair comparison of performance, we evaluate tracers on the same five input datasets per benchmark.
Each dataset contains the test cases generated by fuzzing that benchmark with AFL-QEMU for 24 hours.
Though our results show UnTracer has less than 1\% overhead after one hour of fuzzing, we extend all evaluations to 24 hours to better match previous fuzzing evaluations.

We configure AFL to run with 500ms timeouts and leave all other parameters at their defaults.
We modify AFL so that all non-tracing functionality is removed (e.g., progress reports) and instrument its \texttt{run\_target()} function to collect per-test case timing.
To address noise from the operating system and other sources, we perform eight trials of each dataset.
For each set of trials per dataset, we apply trimmed-mean de-noising~\cite{arnautov_scone:_2016} on each test case's tracing times; the resulting times represent each test case's median tracing performance.

All trials are distributed across two workstations---each with five single-core virtual machines.
Both host systems run Ubuntu 16.04 x86\_64 operating systems, with six-core Intel Core i7-7800X CPU @ 3.50GHz, and 64GB RAM.
All 10 virtual machines run Ubuntu x86\_64 18.04 using VirtualBox. 
We allocate each virtual machine 6GB of RAM.\footnote{Across all trials, we saw no benchmarks exceeding 2GB of RAM usage.}

\subsection{Experiment Infrastructure}
To narrow our focus to tracing overhead, we only record time spent executing/tracing test cases.
To maintain fairness, we run all tracers on the same five pre-generated test case datasets for each benchmark.
For dataset generation, we implement a modified version of AFL that dumps its generated test cases to file.
In our evaluations, we use QEMU as the baseline tracer (since our focus is black-box tracing) to generate the five datasets for each benchmark.

Our second binary---\texttt{TestTrace}---forms the backbone of our evaluation infrastructure.
We implement this using a modified version of AFL---eliminating components irrelevant to tracing (e.g., test case generation and execution monitoring).
Given a benchmark, pre-generated dataset, and tracing mode (i.e., AFL-Clang, AFL-QEMU, AFL-Dyninst, or none (a.k.a. \emph{baseline})), \texttt{TestTrace}: (1) reproduces the dataset's test cases one-by-one, (2) measures  time spent tracing each test case's coverage, and (3) logs each trace time to file.
For UnTracer, we include both the initial full-speed execution and any time spent handling coverage-increasing test cases.

\subsection{Benchmarks}

Our benchmark selection is based on popularity in the fuzzing community and benchmark type.
We first identify candidate benchmarks from several popular fuzzers' trophy cases\footnote{A fuzzer's ``trophy case'' refers to a collection of bugs/vulnerabilities reportedly discovered with that fuzzer.} and public benchmark repositories \cite{zalewski_american_2017, rash_afl-cve:_2017, swiecki_honggfuzz_2018, serebryany_oss-fuzz_2017, google_fuzzer-test-suite:_2018}. 
To maximize benchmark variety, we further partition candidates by their overall type---software \textbf{dev}elopment, \textbf{image} processing, data \textbf{archiv}ing, \textbf{net}work utilities, \textbf{audio} processing, \textbf{doc}ument processing, \textbf{crypto}graphy, and \textbf{web} development. 
After we filter out several candidate benchmarks based on incompatibility with our tracers (e.g., Dyninst-based instrumentation crashes on \texttt{openssl}), we select one benchmark per category: \texttt{bsdtar} (archiv), \texttt{cert-basic} (crypto), \texttt{cjson} (web), \texttt{djpeg} (image), \texttt{pdftohtml} (doc), \texttt{readelf} (dev), \texttt{sfconvert} (audio), and \texttt{tcpdump} (net).

For each benchmark, we measure several other metrics with potential effects on tracing overhead: number of basic blocks; and average number of generated test cases, average rate of coverage-increasing test cases, and average number of 500ms timeouts in 24 hours.
Benchmark basic block totals are computed by enumerating all basic blocks statically using Dyninst \cite{noauthor_dyninst_2018}.
For counting timeouts, we examined the statistics reported by \texttt{afl-fuzz-saveinputs} during dataset generation; using our specified timeout value (500ms), we then averaged the number of timeouts per benchmark among its datasets. 
Lastly, for each benchmark, we counted and averaged the number of test cases generated in all of its 24-hour datasets.

We compile each benchmark using Clang/LLVM, with all compiler options set to their respective benchmark-specific defaults. 
Below, we detail our additional tracer-specific benchmark configurations.

\begin{figure}[!t]
    \centering
    \includegraphics[width=1\columnwidth]{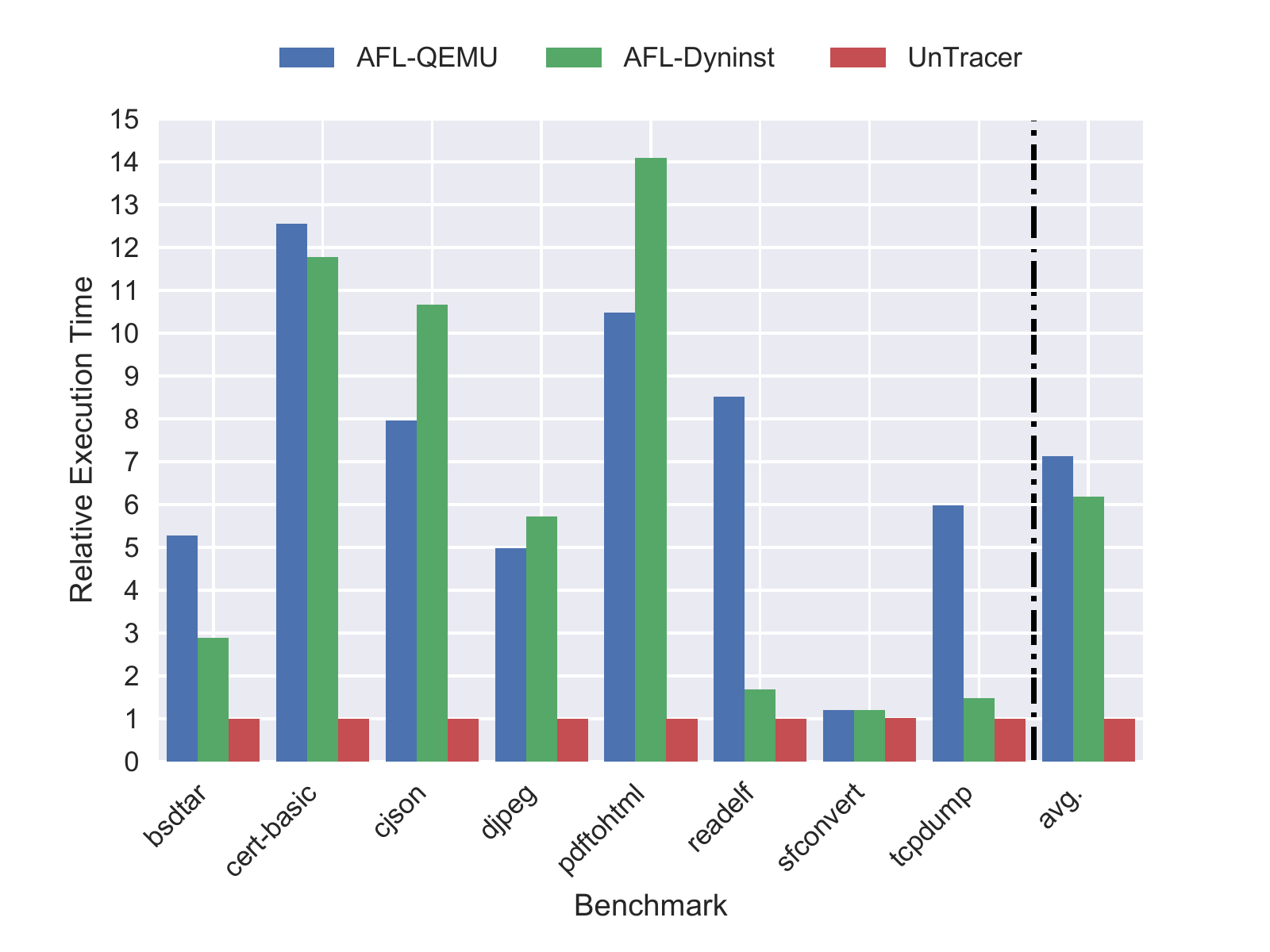}
    \caption{Per-benchmark relative overheads of UnTracer versus black-box binary tracers AFL-QEMU and AFL-Dyninst.}
    \label{fig:tracerOverheadBB}\hrulefill
    \hfill
\end{figure}

\subsubsection{Baseline}
AFL's forkserver-based execution model (also used by UnTracer's interest oracle and tracer binaries) adds a substantial performance improvement over \texttt{execve()}-based execution \cite{zalewski_fuzzing_2014}. 
As each fuzzing tracer in our evaluation leverages forkserver-based execution, we design our ``ground-truth'' benchmark execution models to represent the fastest known execution speeds: a statically-instrumented forkserver without any coverage tracing.
We use a modified copy of AFL's assembler (\texttt{afl-as}) to instrument \emph{baseline} (forkserver-only) benchmark versions.
In each benchmark trial, we use its baseline execution speeds as the basis for comparing each fuzzing tracers' overhead.

\subsubsection{AFL-Clang}
As compiling with AFL-GCC failed for some binaries due to changes in GCC, we instead use AFL-Clang.

\subsubsection{AFL-QEMU}
We only need to provide it the original uninstrumented target binary of each benchmark in our evaluation.

\subsubsection{AFL-Dyninst}
For our AFL-Dyninst evaluations, we instrument each binary using AFL-Dyninst's instrumenter with configuration parameters \texttt{bpatch.setDelayedParsing} set to true; \texttt{bpatch.setLivenessAnalysis} and \texttt{bpatch.setMergeTramp} false; and leave all other configuration parameters at their default settings.

\eat{
    \subsubsection{Intel Processor Trace}
    Intel Processor Trace (IPT) \cite{intel_intel_2017}---a black-box binary hardware-assisted tracer---has seen recent adoption by the fuzzing community \cite{schumilo_kafl:_2017, swiecki_honggfuzz_2018, zhang_ptfuzz:_2018}; however, the lack of any official IPT support in AFL motivated us to implement it ourselves.
    We built a separate version of \texttt{TestTrace}---\texttt{IPT-Tracer}---implementing IPT tracing with Linux's \texttt{perf} subsystem \cite{linux_perf_2018} (similar to honggFuzz \cite{swiecki_honggfuzz_2018}, kAFL \cite{schumilo_kafl:_2017}, and PTfuzz \cite{zhang_ptfuzz:_2018}).
    For coverage tracking, IPT traces must be decoded into their raw basic block information; unfortunately, current decoding solutions introduce significant overhead in fuzzing \cite{swiecki_afl-users_2016}.
    As our aim is to evaluate only IPT's \emph{tracing} performance, we skip trace decoding entirely; thus, our tracing-only implementation (\texttt{IPT-Tracer}) offers a \emph{best-case} estimation of IPT's performance as a black-box binary fuzzing tracer.
    To further isolate IPT's tracing performance, we run \texttt{IPT-Tracer} on each benchmark's assembler-instrumented baseline forkserver version.
}

\subsection{Timeouts}
Coverage tracing is affected by pre-defined execution \emph{timeout} values.
Timeouts act as a ``hard limit''---terminating a test case's tracing if its duration exceeds the timeout's value.
Though timeouts are necessary for halting infinitely-looping test cases, small timeouts prematurely terminate tracing.
For long-running test cases, this results in missed coverage information.
In cases where missed coverage causes coverage-increasing test cases to be misidentified as non-coverage-increasing, this will have cascading effects on test case generation. 
As coverage-guided fuzzers explore the target binary by mutating coverage-increasing test cases, exclusion of timed-out---but otherwise coverage-increasing---test cases results in a higher likelihood of generated test cases being non-coverage-increasing, and thus, slowing coverage indefinitely.

Small timeouts, when hit frequently, distort tracers' overheads, making their performance appear closer to each others'. 
In early experiments with timeouts of 100ms (AFL's default), we observed that, for some datasets, our worst-performing tracers (e.g., AFL-Dyninst, AFL-QEMU) had similar performance to otherwise faster white-box-based tracing (i.e., AFL-Clang).
Upon investigating each tracer's logs, we found that all were timing-out on a significant percentage of the test cases.
This was striking given that the baseline (forkserver-only) benchmark versions had significantly fewer timeouts. 
Thus, a 100ms timeout was too restrictive.
We explored the effect of several different timeout values, with the goal of making each tracer's number of timeouts close to the baseline's (assumed ground truth).

\eat{
    With higher timeout values, we observed that static/dynamic-instrumented binary tracing faced a significantly higher number of timeouts than assembler-instrumented tracing (e.g., AFL-Clang). 
    To identify the \emph{expected} number of timeouts per dataset, we examined the trimmed-mean baseline (forkserver-only) evaluation results, and counted instances where a test case's tracing time matched our execution timeout value.
    After increasing our timeout value to 500ms, we observed a minimal difference in reported timeouts between assembler- and static/dynamic-instrumented tracers, and baseline.
    We first conclude that, if timeouts are set to their optimal values, all tracers' total reported timeouts should be relatively similar to one another. 
    Secondly, timeouts should also be small enough to prevent significant evaluation slowdown resulting from infinite tracing of looping test cases.
    For our evaluations, we feel that a 500ms timeout offers an acceptable balance between coverage completeness and tracing slowdown.
}

\begin{figure}[!t]
    \centering
    \includegraphics[width=1\columnwidth]{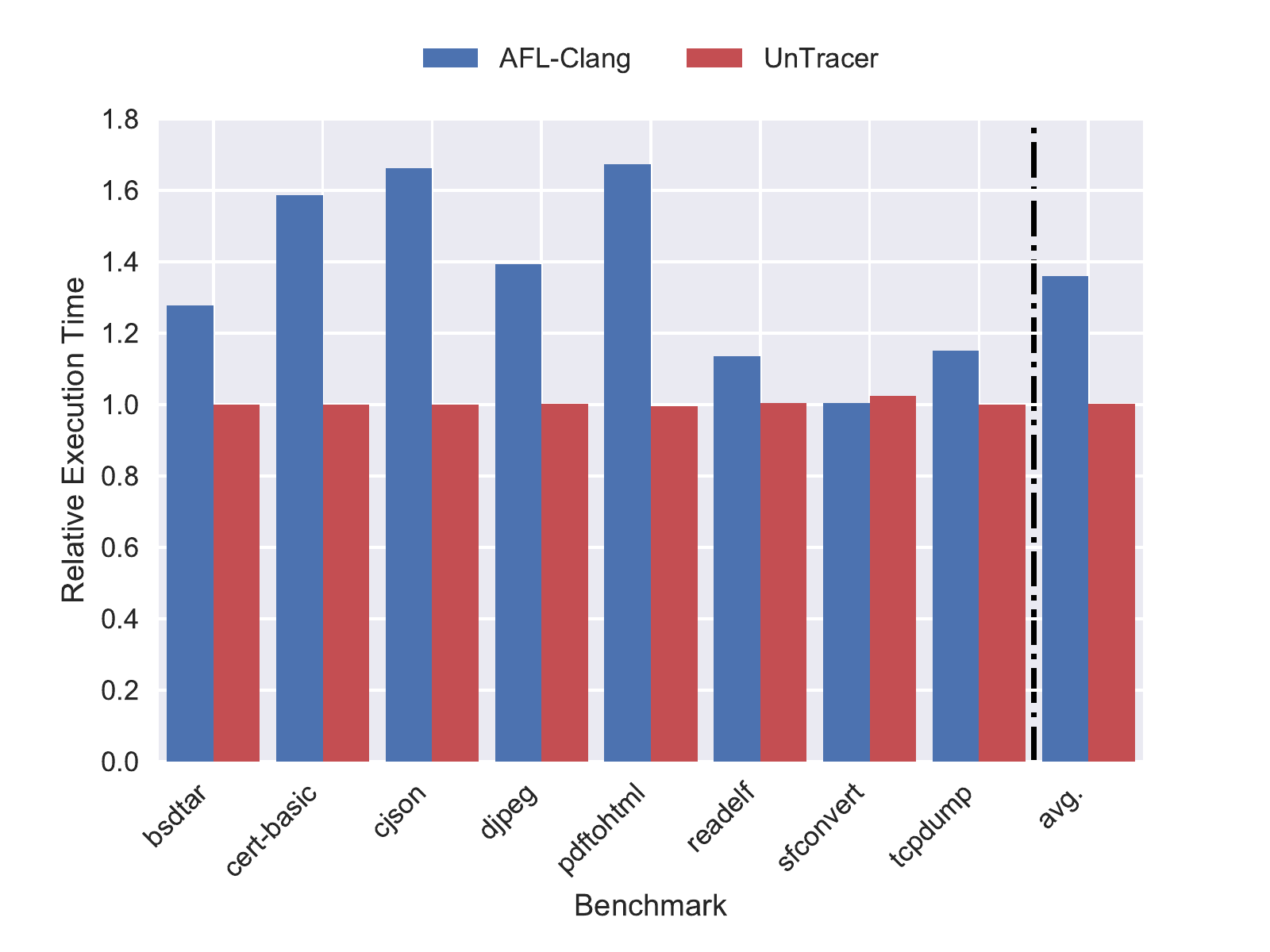}
    \caption{Per-benchmark relative overheads of UnTracer versus white-box binary tracer AFL-Clang.}
    \label{fig:tracerOverheadWB}
    \hrulefill
\end{figure}

\subsection{UnTracer versus Coverage-agnostic Tracing}
\label{sec:eval_overhead}
We examine our evaluation results to identify each fuzzing tracer's overhead per benchmark.
For each tracer's set of trials per benchmark dataset, we employ trimmed-mean de-noising (shown to better reveal median tendency \cite{arnautov_scone:_2016}) at test case level---removing the top and bottom 33\% outliers---to reduce impact of system interference on execution speeds.
We then take the resulting five trimmed-mean dataset overheads for each tracer-benchmark combination and average them to obtain tracer-benchmark overheads.
Lastly, we convert all averaged tracer-benchmark overheads to \emph{relative execution times} with respect to baseline (e.g., a relative execution time of 1.5 equates to 50\% overhead).

In the following sections, we compare the performance of UnTracer to three popular \emph{coverage-agnostic} tracing approaches.
We first explore the performance of two black-box binary fuzzing tracers: AFL-QEMU (dynamic) and AFL-Dyninst (static).
Secondly, we compare UnTracer's performance against that of the white-box binary fuzzing tracer AFL-Clang (static assembler-instrumented tracing).

\begin{figure}[!t]
     \centering
     \includegraphics[width=1\columnwidth]{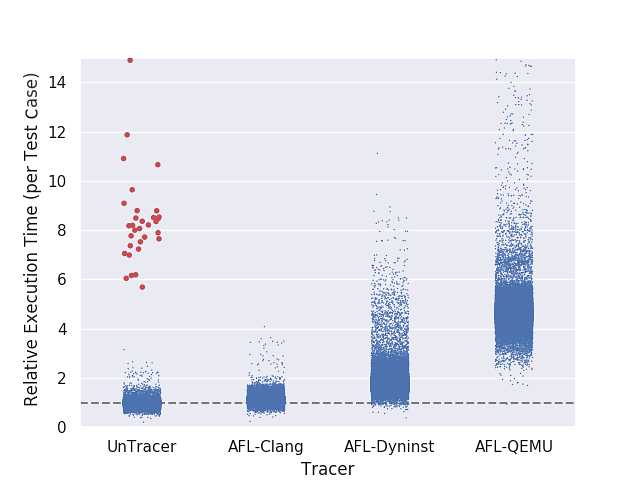}
     \caption{Distribution of each tracer's relative execution time averaged per-test case for one 24-hour \texttt{cjson} dataset. The horizontal grey dashed line represents the average baseline execution speed. Red dots represent coverage-increasing test cases identified by UnTracer.}
     \label{fig:swarm}
     \hrulefill
\end{figure}

\subsubsection{Black-box binary tracing}
\label{sec:eval:black-box-bin-tracing}
As shown in Figure~\ref{fig:tracerOverheadBB}, we compare UnTracer's performance to two popular black-box binary fuzzing tracers---AFL's dynamically-instrumented tracing via QEMU user-mode emulation (AFL-QEMU) \cite{zalewski_afl-users_2016}, and Dyninst-based static binary rewriting-instrumented tracing (AFL-Dyninst) \cite{talos-vulndev_afl-dyninst_2018}.
For one benchmark (\texttt{sfconvert}), AFL-QEMU and AFL-Dyninst have similar relative execution times (1.2 and 1.22, respectively) to UnTracer (1.0); however; by looking at the different datasets for \texttt{sfconvert}, we observe a clear trend between higher number of timeouts and lower tracing overheads across all tracers (Table~\ref{tab:binaries}).
In our evaluations, a 500ms test case timeout significantly overshadows a typical test case execution of 0.1--1.0ms.

AFL-Dyninst outperforms AFL-QEMU in three benchmarks (\texttt{bsdtar, readelf, tcpdump}), but as these benchmarks all vary in complexity (e.g., number of basic blocks, execution times, etc.), we are unable to identify which benchmark characteristics are optimal for AFL-Dyninst's performance.
Across all benchmarks, UnTracer achieves an average relative execution time of 1.003 (0.3\% overhead), while AFL-QEMU and AFL-Dyninst average relative execution times of 7.12 (612\% overhead) and 6.18 (518\% overhead), respectively.
The average Relative Standard Deviation (RSD) for each tracer was less than 4\%.
In general, our results show UnTracer reduces the overhead of tracing black-box binaries by up to four orders of magnitude.

\par{\textbf{Mann Whitney U-test scoring:}}
Following Klees et al.'s~\cite{klees_evaluating_2018} recommendation, we utilize the Mann Whitney U-test to determine if UnTracer's execution overhead is stochastically smaller than AFL-QEMU's and AFL-Dyninst's. 
First we compute all per-dataset execution times for each benchmark\footnote{We ignore \texttt{sfconvert} in all statistical evaluations as its high number of timeouts results in all tracers having similar overhead.} and tracer combination; then for each benchmark dataset we apply the Mann Whitney U-test with 0.05 significance level on execution times of UnTracer versus AFL-QEMU and UnTracer versus AFL-Dyninst.
Averaging the resulting $p$-values for each benchmark and tracer combination is less than .0005 for UnTracer compared (pair-wise) to AFL-QEMU and AFL-Dyninst. Given that these $p$-values are much smaller than the 0.05 significance level, we conclude there exists a statistically significant difference in the median execution times of UnTracer versus AFL-QEMU and AFL-Dyninst.

\par{\textbf{Vargha and Delaney $\hat{A}_{12}$ scoring:}}
To determine the \emph{extent} to which UnTracer's execution time outperforms AFL-QEMU's and AFL-Dyninst's, we apply Vargha and Delaney's $\hat{A}_{12}$ statistical test~\cite{vargha_critique_2000}.
For all comparisons among benchmark trials the resulting $\hat{A}_{12}$ statistic is 1.0---exceeding the conventionally large effect size of 0.71.
Thus we conclude that the difference in execution times between UnTracer versus either black-box tracer is statistically large.

\eat{
    In Figure~\ref{fig:eval_benchmarks_ipt}, we evaluate UnTracer versus an emerging black-box binary hardware-assisted fuzzing tracing approach---Intel Processor Trace (IPT) \cite{intel_intel_2017}.
    \textbf{TODO: IPT results}.
}

\subsubsection{White-box binary tracing}
In Figure~\ref{fig:tracerOverheadWB}, we show the benchmark overheads of UnTracer, and AFL's white-box binary (static assembly-time instrumented) tracer AFL-Clang.
AFL-Clang averages a relative execution time of 1.36 (36\% overhead) across all eight benchmarks, while UnTracer averages 1.003 (0.3\% overhead) (average RSD for each tracer was less than 4\%).
As is the case for black-box binary tracers AFL-QEMU and AFL-Dyninst, in one benchmark with a large number of timeouts---\texttt{sfconvert}---AFL-Clang's performance is closest to baseline (nearly matching UnTracer's).

\par{\textbf{Mann Whitney U-test scoring:}}
On average per dataset, the resulting $p$-values ranged from .00047 to .015---though only in one instance did the $p$-value exceed .0005. 
Thus we conclude that there is a statistically significant difference in median execution times of UnTracer versus AFL-Clang. 

\par{\textbf{Vargha and Delaney $\hat{A}_{12}$ scoring:}}
Among all trials the resulting $\hat{A}_{12}$ statistics range from 0.76 to 1.0. As the minimum of this range exceeds 0.71, we conclude UnTracer's execution time convincingly outperforms AFL-Clang's.

Figure~\ref{fig:swarm} shows the distributions of overheads for each tracer on one dataset of the \texttt{cjson} benchmark. 
The coverage-increasing test cases (red dots) are clearly separable from the non-coverage-increasing test cases for UnTracer, with the coverage-increasing test cases incurring double the overhead of tracing with AFL-Dyninst alone.

\begin{figure}[!t]
     \centering
     \includegraphics[width=1\columnwidth]{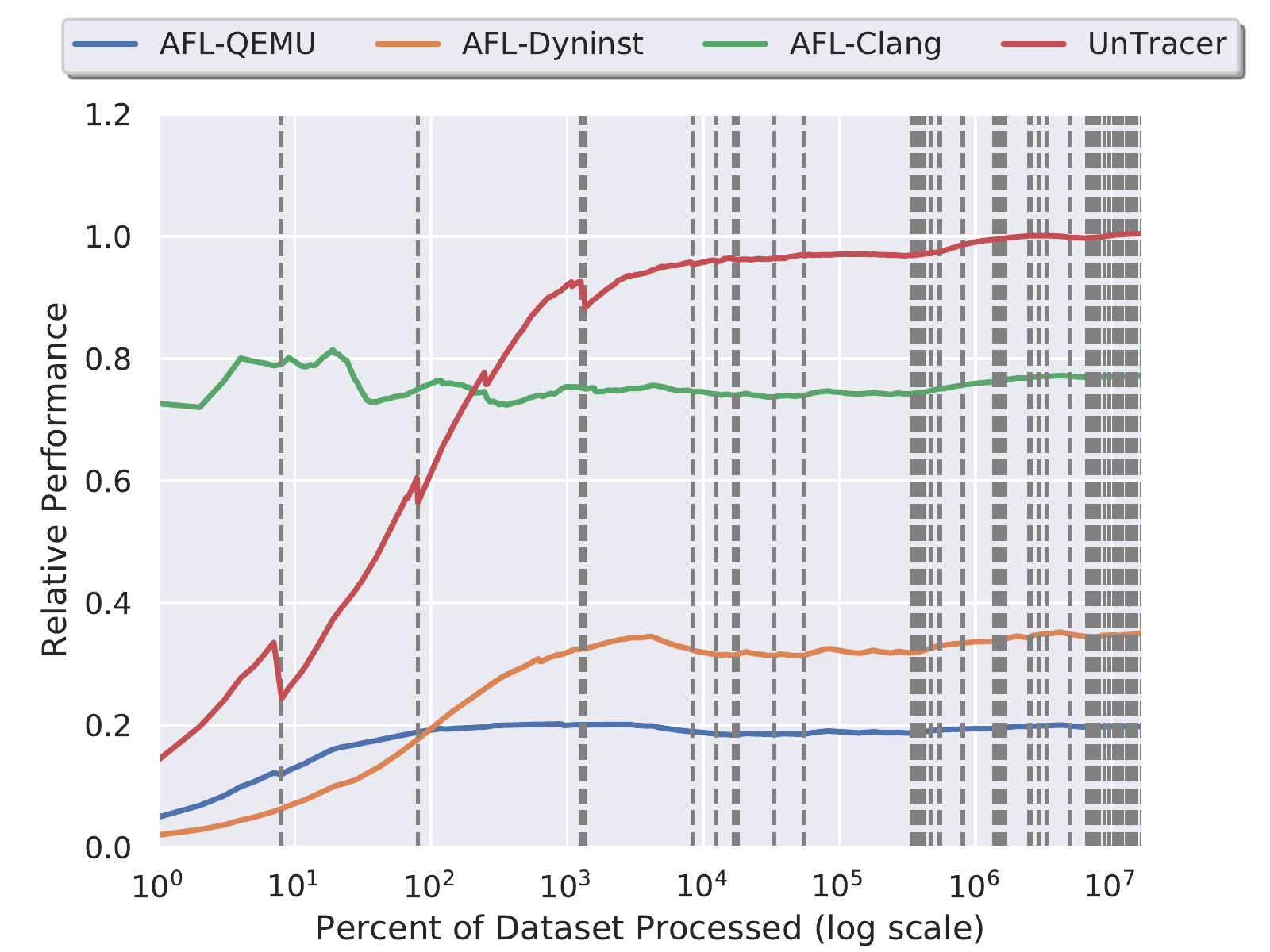}
     \caption{Averaged relative performance of all tracers over the percentage of test cases processed for one 24-hour \texttt{bsdtar} dataset. Here, 1.0 refers to baseline (maximum) performance. Each grey dashed vertical line represents a coverage-increasing test case.}
     \label{fig:overheadOverTime}
     \hrulefill
\end{figure}

Figure~\ref{fig:overheadOverTime} shows how UnTracer's overhead evolves over time and coverage-increasing test cases. Very early in the fuzzing process, the rate of coverage-increasing test cases is high enough to degrade UnTracer's performance.  
As time progresses, the impact of a single coverage-increasing test case is inconsequential and UnTracer gradually approaches 0\% overhead. 
In fact, by 1000 test cases, UnTracer has 90\% of the native binary's performance. 
This result also shows that there is an opportunity for a hybrid coverage-guided tracing model, where initial test cases are always traced until the rate of coverage-increasing test cases diminishes to the point where UnTracer becomes beneficial.

\begin{figure}
     \centering
     \includegraphics[width=1\columnwidth]{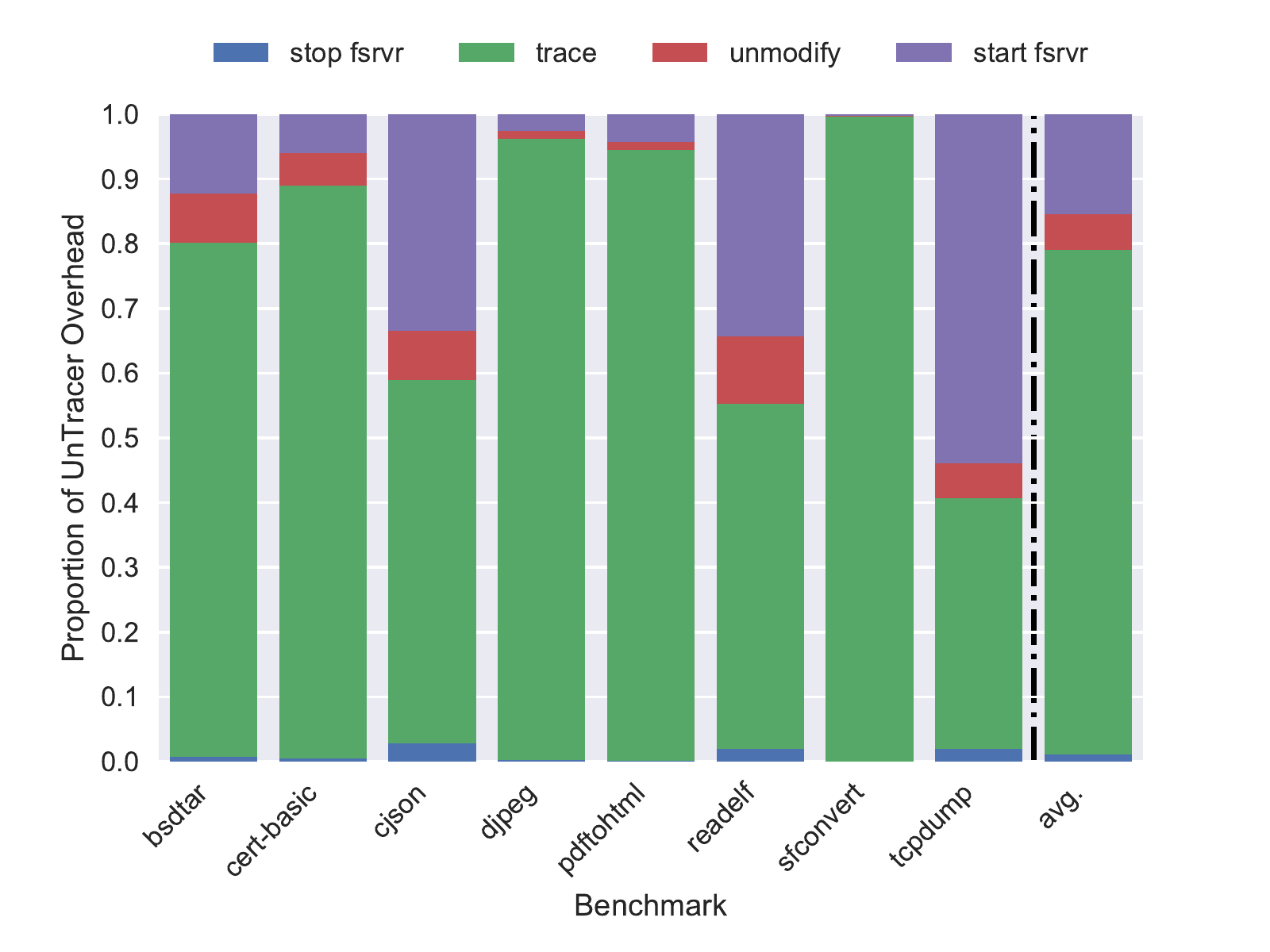}
     \caption{Visualization of the overheads per UnTracer's four components related to coverage-increasing test case processing for each benchmark.}
     \label{fig:untracerOverhead}
     \hrulefill
\end{figure}

\subsection{Dissecting UnTracer's Overhead}

While Untracer achieves significantly lower overhead compared to conventional coverage-agnostic tracers (i.e., AFL-QEMU, AFL-Dyninst, AFL-Clang), it remains unclear which operations are the most performance-taxing.
As shown in Algorithm~\ref{alg:untracer}, UnTracer's high-level workflow comprises the following: (1) starting the interest oracle and tracer binary forkservers; (2) identifying coverage-increasing test cases by executing them on the oracle; (3) tracing coverage-increasing test cases' code coverage by executing them on the tracer; (4) stopping the oracle's forkserver; (5) unmodifying (removing interrupts from) basic blocks in the oracle; and (6) restarting the oracle's forkserver.
Since UnTracer identifies coverage-increasing test cases as those which trigger the oracle's interrupt, non-coverage-increasing test cases---the overwhelming majority---exit the oracle cleanly without triggering any interrupts.
Thus, executing non-coverage-increasing test cases on the oracle is equivalent to executing them on the original (baseline) binary.
Based on this, UnTracer's only overhead is due to processing coverage-increasing test cases. 

In our evaluation of UnTracer's overhead, we add timing code around each component run for every coverage-increasing test case: coverage tracing with the tracer (\texttt{trace}), stopping the oracle's forkserver (\texttt{stop fsrvr}), unmodifying the oracle (\texttt{unmodify}), and restarting the oracle (\texttt{start fsrvr}).
We average all components' measured execution times across all coverage-increasing test cases, and calculate their respective proportions of UnTracer's total overhead.
Figure~\ref{fig:untracerOverhead} shows the breakdown of all four components' execution time relative to total overhead.
The graph shows that the two largest components of UnTracer's overhead are coverage tracing and forkserver restarting.

\par{\textbf{Tracing:}}
Unsurprisingly, coverage tracing (\texttt{trace}) contributes to the almost 80\% of UnTracer's overhead across all benchmarks.
Our implementation relies on Dyninst-based static binary rewriting-instrumented black-box binary tracing.
As our evaluation results (Figure~\ref{fig:tracerOverheadBB}) show, in most cases, Dyninst adds a significant amount of overhead.
Given UnTracer's compatibility with other binary tracers, there is an opportunity to take advantage of faster tracing (e.g., AFL-Clang in a white-box binary tracing scenario) to lower UnTracer's total overhead.

\par{\textbf{Forkserver restarting:}}
Restarting the oracle's forkserver (\texttt{start fsrvr}) is the component with second-highest overhead.
In binaries with shorter test case execution times (e.g., \texttt{cjson, readelf}, and \texttt{tcpdump}), the proportion of tracing time decreases, causing more overhead to be spent on forkserver restarting.
Additionally, in comparison to UnTracer's constant-time forkserver-\emph{stopping} operation (\texttt{stop fsrvr}), forkserver-restarting relies on costly process creation (e.g., \texttt{fork(), execve()}) and inter-process communication (e.g., \texttt{pipe(), read(), write()}).
Previous work looks at optimizing these system calls for fuzzing \cite{xu_designing_2017}, but given UnTracer's low overhead in our evaluation, further optimization adds little performance improvement.
However, we can imagine niche contexts where such approaches would yield meaningful performance improvements.

\subsection{Overhead versus Rate of Coverage-increasing test cases}

\begin{figure}
     \centering
     \includegraphics[width=\columnwidth]{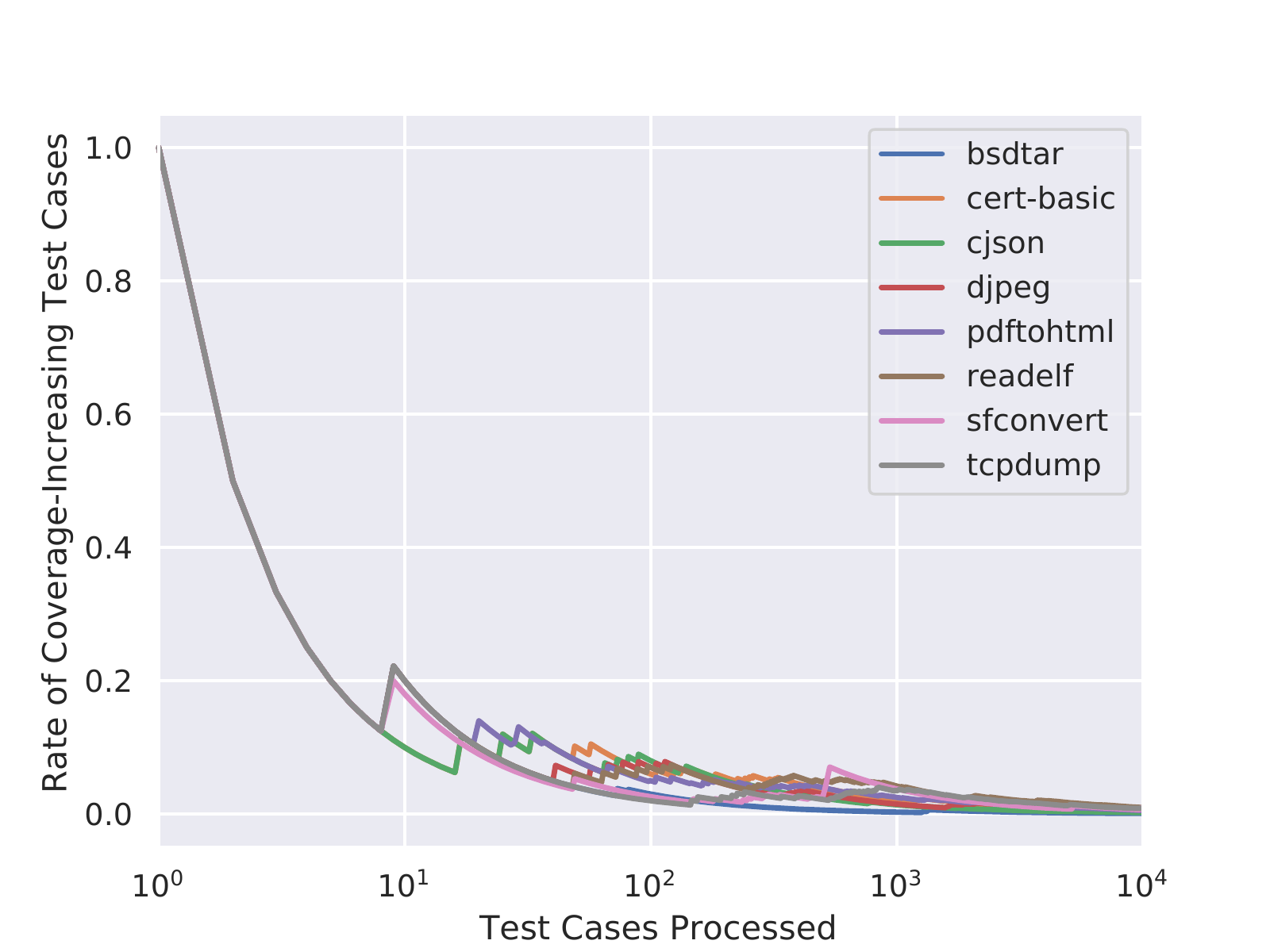}
     \caption{The rates of coverage-increasing test cases encountered over the total number of test cases processed, per benchmark.}
     \label{fig:interestingRate}
     \hrulefill
\end{figure}

Below, we discuss the potential performance advantage of a hybrid approach combining coverage-guided and coverage-agnostic tracing (e.g., AFL \cite{zalewski_american_2017}, libFuzzer \cite{serebryany_continuous_2016}, honggFuzz \cite{swiecki_honggfuzz_2018}).
In contrast to existing fuzzing tracers, which face high overhead due to tracing \emph{all} generated test cases, UnTracer achieves near-zero overhead by tracing only \emph{coverage-increasing} test cases---the rate of which decreases over time for all benchmarks (Figure~\ref{fig:interestingRate}).
Compared to AFL, UnTracer's coverage tracing is slower on average---largely due to its trace reading/writing relying on slow file input/output operations.
Thus, as is the case in our evaluations (Table~\ref{tab:binaries}), coverage-guided tracing offers significant performance gains when \emph{few} generated test cases are coverage-increasing.
For scenarios where a higher percentage of test cases are coverage-increasing (e.g., fuzzers with ``smarter'' test case generation \cite{rawat_vuzzer:_2017, chen_angora:_2018, li_steelix:_2017}), our approach may yield less benefit.

In such cases, overhead may be minimized using a \emph{hybrid} fuzzing approach that switches between coverage-guided and coverage-agnostic tracing, based on the observed rate of coverage-increasing test cases.
We first identify a \emph{crossover threshold}---the rate of coverage-increasing test cases at which coverage-guided tracing's overhead exceeds coverage-agnostic tracing's.
During fuzzing, if the rate of coverage-increasing test cases drops below the threshold, coverage-guided tracing becomes the optimal tracing approach; its only overhead is from tracing the few coverage-increasing test cases.
Conversely, if the rate of coverage-increasing test cases exceeds the threshold, coverage-agnostic tracing (e.g., AFL-Clang, AFL-QEMU, AFL-Dyninst) is optimal.

\begin{figure}
    \centering
    \includegraphics[width=1\columnwidth]{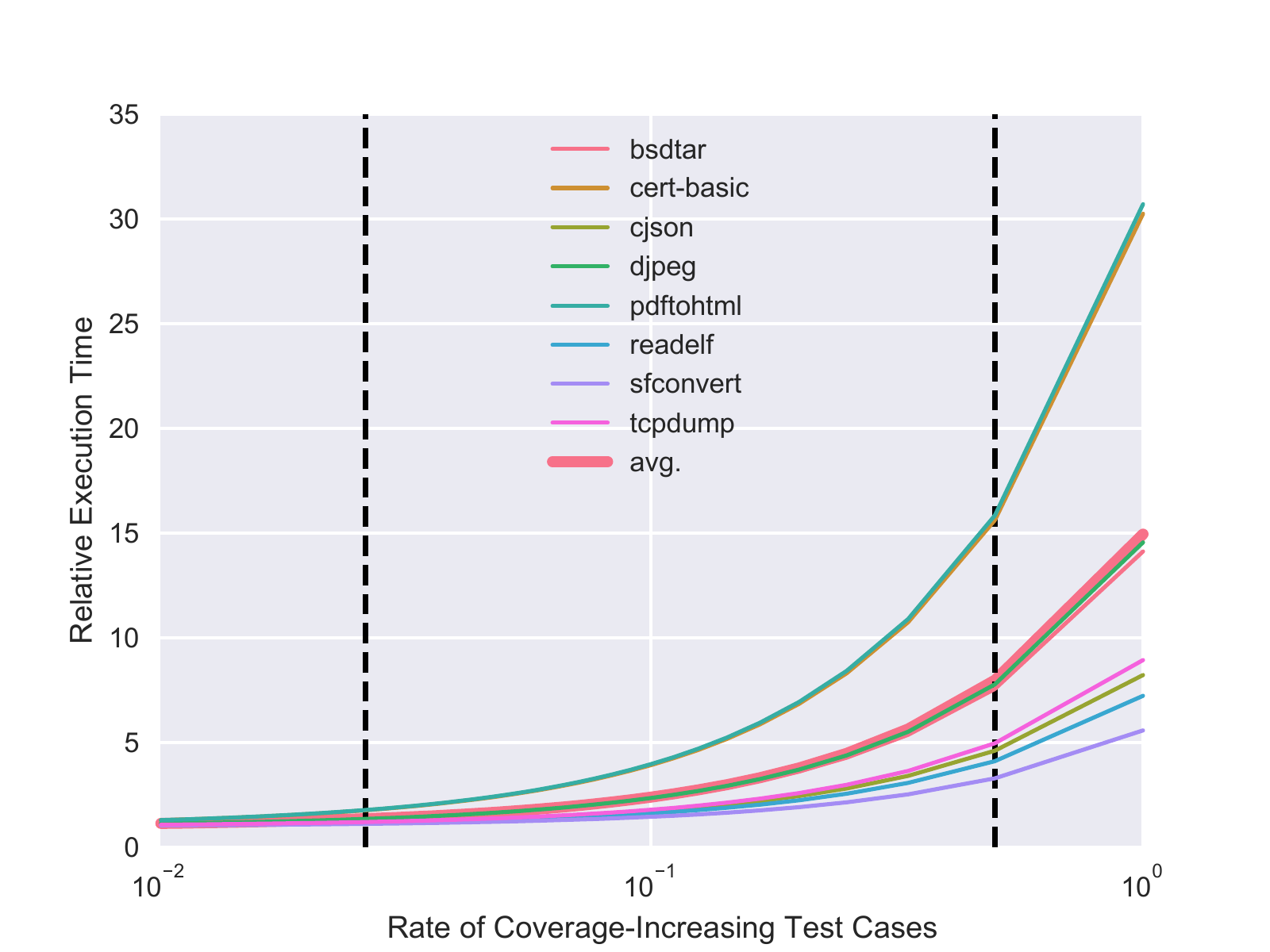}
    \caption{Model of the relationship between coverage-increasing test case rate and UnTracer's overhead per test case. For all rates left of the leftmost dashed vertical line, UnTracer's overhead per test case is less than AFL-Clang's. Likewise, for all rates left of the rightmost dashed vertical line, it is less than AFL-QEMU's and AFL-Dyninst's. Not shown is the average rate of coverage-increasing test cases observed during our evaluations (4.92E-3).}
    \label{fig:untracerOverheadsSweep}
    \hrulefill
\end{figure}

To develop a universally-applicable threshold for all tracing approaches, we average the overheads of coverage-increasing test cases across all trials in our tracer-benchmark evaluations.
We then model overhead as a function of the rate of coverage-increasing test cases; we apply this model to identify the coverage-increasing test case rates where UnTracer's overhead exceeds AFL-Clang's, and AFL-QEMU's and AFL-Dyninst's.
As shown in Figure~\ref{fig:untracerOverheadsSweep}, for all rates of coverage-increasing test cases below 2\% (the leftmost dashed vertical line), UnTracer's overhead per test case is less than AFL-Clang's.
Similarly, UnTracer's overhead per test case is less than AFL-QEMU's and AFL-Dyninst's for all rates less than 50\% (the rightmost vertical dashed line).

\makeatletter{}\section{Hybrid Fuzzing Evaluation}
\label{sec:evaluation_fuzzing}

State-of-the-art hybrid fuzzers (e.g., Driller~\cite{stephens_driller:_2016} and QSYM~\cite{yun_qsym:_2018}) combine program-directed mutation (e.g., via concolic execution) with traditional blind mutation (e.g., AFL~\cite{zalewski_american_2017}). Hybrid approaches offer significant gains in code coverage at the cost of reduced test case execution rate. In this section, we compare UnTracer, Clang~\cite{zalewski_american_2017} (white-box tracing), and QEMU~\cite{zalewski_american_2017} (black-box dynamically-instrumented tracing) implementations of the state-of-the-art hybrid fuzzer QSYM on seven of our eight benchmarks.\footnote{We exclude \texttt{sfconvert} from this evaluation since the QEMU-based variant of QSYM crashes on all eight experimental trials.} Exploring the benefit of UnTracer in a hybrid fuzzing scenario is important as hybrid fuzzers make a fundamental choice to spend less time executing test cases (hence tracing) and more time on mutation. While we provide an estimate of the impact hybrid fuzzing has on coverage-guided tracing's value in Section~\ref{sec:perf}, this section provides concrete data on the impact to UnTracer of a recent hybrid fuzzer.

\subsubsection{Implementing QSYM-UnTracer}

We implemented~\cite{nagy_untracer-afl:_2019} QSYM-UnTracer in QSYM's core AFL-based fuzzer, which tracks coverage (invoked by \texttt{run\_target()}) in several contexts: test case trimming (\texttt{trim\_case()}), test case calibration (\texttt{calibrate\_case()}), test case saving (\texttt{save\_if\_interesting()}), hybrid fuzzing syncing (\texttt{sync\_fuzzers()}), and the ``common'' context used for most test cases (\texttt{common\_fuzz\_stuff()}). 
Below we briefly discuss design choices specific to each.

\par{\textbf{Trimming and calibration:}} test case trimming and calibration must be able to identify changes in a priori coverage. 
Thus the interest oracle is unsuitable since it only identifies \emph{new} coverage, and we instead utilize only the tracer binary. 

\par{\textbf{Saving timeouts:}} A sub-procedure of test case saving involves identifying unique timeout-producing and unique hang-producing test cases by tracing and comparing their coverage to a global timeout coverage.
Since AFL only tracks this information for reporting purposes (i.e., timeouts and hangs are not queued), and using an interest oracle or tracer would ultimately add unwanted overhead for binaries with many timeouts (e.g., \texttt{djpeg} (Table~\ref{tab:binaries})), we configure UnTracer-AFL, AFL-Clang, and AFL-QEMU to only track \emph{total} timeouts.

For all other coverage contexts we implement the UnTracer interest oracle and tracer execution model as described in Section~\ref{sec:implementation}.

\begin{figure}[!t]
    \centering
    \includegraphics[width=1\columnwidth]{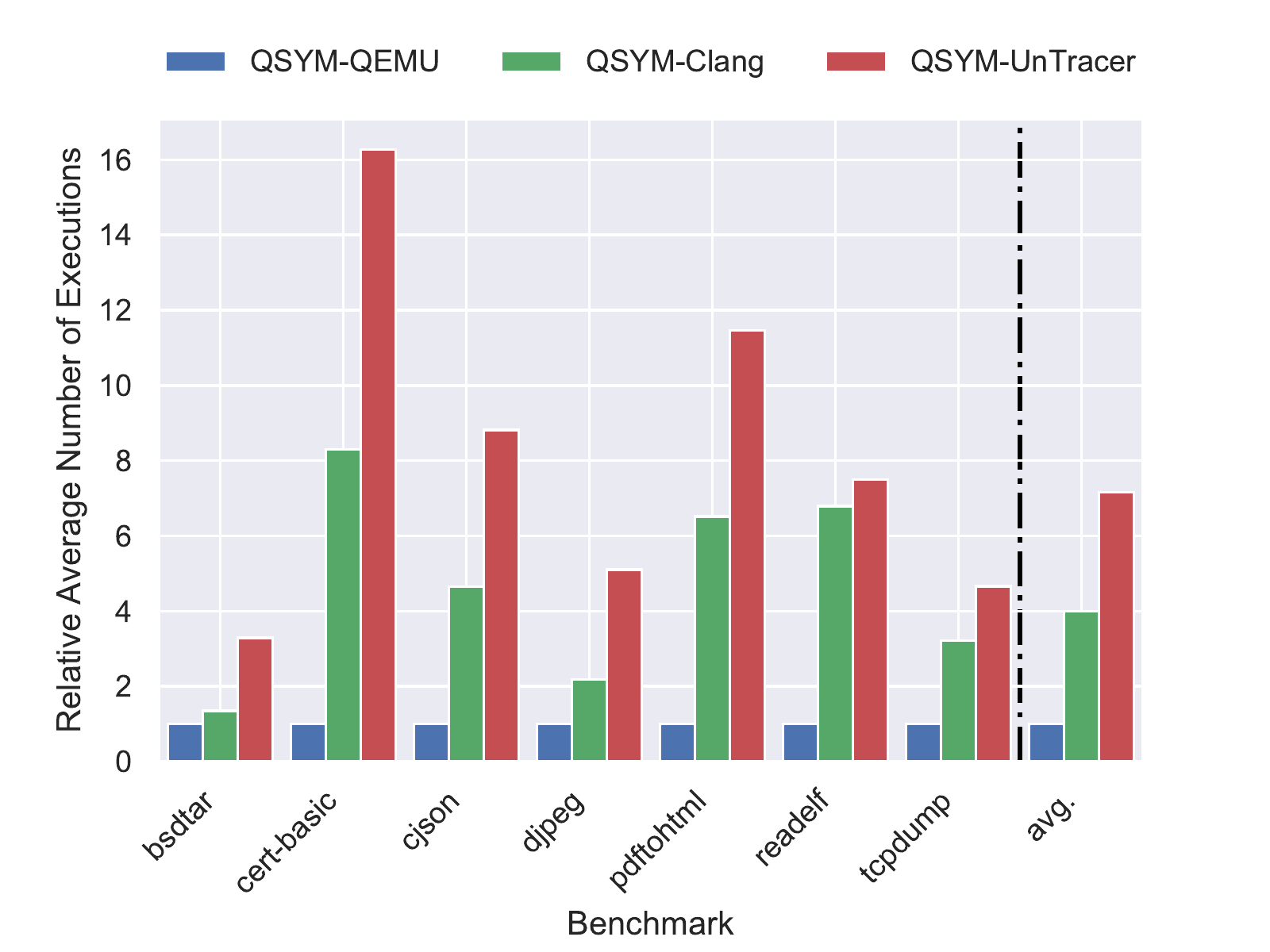}
    \caption{Per-benchmark relative average executions in 24 hours of QSYM-UnTracer versus QSYM-QEMU and QSYM-Clang.}
    \label{fig_fullFuzzerEval}
    \hrulefill
\end{figure}

\subsection{Evaluation Overview}

To identify the performance impact from using UnTracer in hybrid fuzzing we incorporate it in the state-of-the-art hybrid fuzzer QSYM and evaluate its against existing Clang-~\cite{zalewski_american_2017} and QEMU-based~\cite{zalewski_american_2017} QSYM implementations.
Our experiments compare the number of test cases executed for all three hybrid fuzzer variants for seven of the eight benchmarks from Section~\ref{sec:evaluation_tracing} (Table~\ref{tab:binaries}) with 100ms timeouts. To account for randomness, we average the number of test cases executed from 8, 24-hour trials for each variant/benchmark combination. 
To form an average result for each variant across all benchmarks, we compute a per-variant geometric mean.

We distribute all trials across eight virtual machines among four workstations.
Each host is a six-core Intel Core i7-7800X CPU @ 3.50GHz with 64GB of RAM that runs two, two-CPU 6GB virtual machines.
All eight virtual machines run Ubuntu 16.04 x86\_64 (as opposed to 18.04 for previous experiments due to QSYM requirements).
Figure~\ref{fig_fullFuzzerEval} presents the results for each benchmark and the geometric mean across all benchmarks scaled to our baseline of the number of test cases executed by QSYM-QEMU.

\subsection{Performance of UnTracer-based Hybrid Fuzzing}

As shown in Figure~\ref{fig_fullFuzzerEval}, on average, QSYM-UnTracer achieves 616\% and 79\% more test case executions than QSYM-QEMU and QSYM-Clang, respectively.
A potential problem we considered was the overhead resulting from excessive test case trimming and calibration.
Since our implementation of QSYM-UnTracer defaults to the slow tracer binary for test case trimming and calibration, an initial problem we considered was the potential overhead resulting from either operation. 
However, our results show that the performance advantage of interest oracle-based execution (i.e., the ``common case'') far outweighs the performance deficit from trimming and calibration tracing.

\makeatletter{}\section{Discussion}

Here we consider several topics related to our evaluation and implementation. 
First, we discuss the emergence of hardware-assisted coverage tracing, offering a literature-based estimation of its performance with and without coverage-guided tracing.
Second, we detail the modifications required to add basic block edge coverage support to UnTracer and the likely performance impact of moving to edge-based coverage.
Lastly, we highlight the engineering needed to make UnTracer fully support black-box binaries.

\subsection{UnTracer and Intel Processor Trace}

Recent work proposes leveraging hardware support for more efficient coverage tracing. kAFL~\cite{schumilo_kafl:_2017}, PTfuzz~\cite{zhang_ptfuzz:_2018}, and honggFuzz~\cite{swiecki_honggfuzz_2018} adapt Intel Processor Trace (IPT)~\cite{intel_intel_2017} for black-box binary coverage tracing. 
IPT saves the control-flow behavior of a program to a reserved portion of memory as it executes. 
After execution, the log of control-flow information is used in conjunction with an abstract version of the program to generate coverage information. 
Because monitoring occurs at the hardware-level, it is possible to completely capture a program's dynamic coverage at the basic block, edge, or path level incurring modest run time overheads. 
The three main limitations of IPT are its requirement of a supporting processor, time-consuming control-flow log decoding, and its compatability with only x86 binaries.

Despite these limitations, it is important to understand how IPT impacts coverage-guided tracing. 
From a high level, coverage-guided tracing works with IPT because it is orthogonal to the tracing mechanism. 
Thus, an IPT variant of UnTracer would approach 0\% overhead sooner than our Dyninst-based implementation due to IPT's much lower tracing overhead. 
From a lower level, the question arises as to the value of coverage-guided tracing with relatively cheap black-box binary coverage tracing. 
To estimate IPT's overhead in the context of our evaluation, we look to previous work. 
Zhang et al.~\cite{zhang_ptfuzz:_2018} present a fuzzing-oriented analysis of IPT that shows it averaging around 7\% overhead relative to AFL-Clang-fast. 
Although we cannot use this overhead result directly as we compile all benchmarks with AFL-Clang, according to AFL's author, AFL-Clang is 10--100\% slower than AFL-Clang-fast~\cite{zalewski_american_2017}. 
By applying these overheads to the average overhead of 36\% of AFL-Clang from our evaluation, AFL-Clang-fast's projected overhead is between 18--32\% and IPT's projected overhead is between 19--35\%.

\eat{
    \begin{table*}
    \centering
    \small
    \begin{tabular}{ l c c c c c c c c c }
         & \texttt{bsdtar} & \texttt{cert-basic} & \texttt{cjson} & \texttt{djpeg} & \texttt{pdftohtml} & \texttt{readelf} & \texttt{sfconvert} & \texttt{tcpdump} & \textbf{avg.} \\ 
    \hline
    \hline
    \textbf{PCNT} & 15\% & 14\% & 2\% & 6\% & 0\% & 0\% & 11\% & 6\% & 7\% \\
    \end{tabular}
    \vspace{0.2cm}
    \caption{CAPTION
    }
    \label{tab:binaries:criticaledges}
    \hrulefill{}
    \end{table*}
}
\eat{
\begin{table}
\centering
\small
\begin{tabular}{ l c }
\textbf{Benchmark} & \textbf{Critical Edges} \\
\hline
\hline
\texttt{bsdtar} & 14.83\%\\ 
\texttt{cert-basic} & 14.17\%\\
\texttt{cjson} & 2.42\%\\
\texttt{djpeg} & 5.65\%\\
\texttt{pdftohtml} & 0\%\\
\texttt{readelf} & 0\%\\
\texttt{sfconvert} & 10.59\% \\
\texttt{tcpdump} & 6.31\% \\
\textbf{average} & 6.74\% \\
\end{tabular}
\vspace{0.2cm}
\caption{Percentages of critical edges per benchmark.
}
\label{tab:binaries:criticaledges}
\hrulefill{}
\end{table}
}

\subsection{Incorporating Edge Coverage Tracking}
\label{sec:discuss_edge}

\begin{figure}[!t]
    \centering
    \frame{\includegraphics[width=1.0\linewidth]{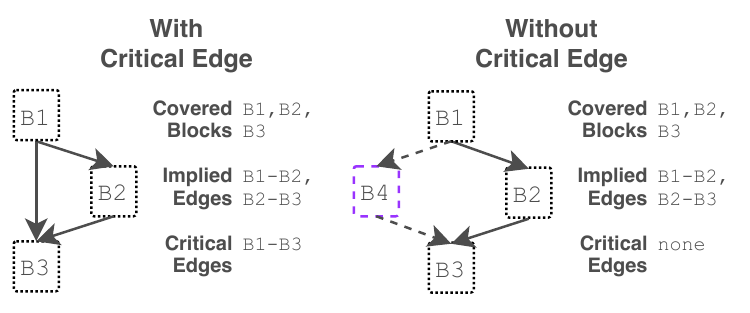}}
    \caption{An example of the \emph{critical edge problem} (left) and its solution (right). To remove the critical edge \texttt{B1-B3}, an empty ``dummy'' block (\texttt{B4}) is inserted to introduce two new edges, \texttt{B1-B4} and \texttt{B4-B3}. Such approach is widely used by software compilers to optimize flow analyses \cite{muchnick_advanced_1997}.}
    \label{fig:criticaledges}
    \hrulefill
\end{figure}

As discussed in Section~\ref{sec:background:cggbf}, two coverage metrics dominate the fuzzing literature: basic blocks and basic block edges.
UnTracer, our implementation of coverage-guided tracing, uses basic block coverage.
Alternatively, many popular fuzzers (e.g., AFL~\cite{zalewski_american_2017}, libFuzzer~\cite{serebryany_continuous_2016}, honggFuzz~\cite{swiecki_honggfuzz_2018}) use edge coverage.
While the trade-offs between basic block and edge coverage metrics have yet to be studied with respect to fuzzing outcomes, we believe that it is important to consider coverage-guided tracing's applicability to edge coverage metrics.

The first point to understand is that most fuzzers that use edge coverage metrics actually rely on basic block-level tracing~\cite{noauthor_sanitizercoverage:_2018}. 
Key to enabling accurate edge coverage while only tracing basic blocks is the removal of critical edges. 
A critical edge is an edge in the control-flow graph whose starting/ending basic blocks have multiple outgoing/incoming edges, respectively~\cite{muchnick_advanced_1997}. 
Critical edges make it impossible to identify which edges are covered from knowing only the basic blocks seen during execution.
This inflates coverage and causes the fuzzer to erroneously discard coverage-increasing inputs.

The solution to the critical edge problem is to split each by inserting an intermediate basic block, as shown in Figure~\ref{fig:criticaledges}. 
The inserted ``dummy'' basic block consists of a direct control-flow transfer to the original destination basic block. 
For white-box binaries, edge-tracking fuzzers honggFuzz~\cite{swiecki_honggfuzz_2018} and libFuzzer~\cite{serebryany_continuous_2016} fix critical edges during compilation~\cite{noauthor_sanitizercoverage:_2018}.
This approach works for white-box use cases of coverage-guided tracing as well.
Unfortunately, how to adapt this approach to black-box binaries is an open technical challenge.

With respect to performance, the impact of moving from basic block coverage to edge coverage is less clear. 
It is clear that, given that edge coverage is a super-set of basic block coverage, the rate of coverage-increasing test cases will increase. 
To determine if the increase in the rate of coverage-increasing test cases is significant enough to disrupt the asymmetry that gives coverage-guided tracing its performance advantage, we reference the results in Figure~\ref{fig:untracerOverheadsSweep} and Table~\ref{tab:perfresults2}. 
Given that seven out of eight of our benchmarks have rates of coverage-increasing test cases below 1 in 100,000 and Figure~\ref{fig:untracerOverheadsSweep} shows that UnTracer provides benefit for rates below 1 in 50, moving to edge-based coverage needs to induce a 4-orders-of-magnitude increase in the rate of coverage-increasing test cases to undermine UnTracer's value.  
Such an increase is unlikely given Table~\ref{tab:perfresults2}, which shows that even for fuzzers using edge coverage, the rate of coverage-increasing test cases is in line with the rates in our evaluation. 
Thus, given UnTracer's near-0\% overhead, we expect that any increase in the rate of coverage-increasing test cases due to moving to edge coverage will \emph{not} change the high-level result of this paper.

\subsection{Comprehensive Black-Box Binary Support}

Niche fuzzing efforts desire support for black-box (source-unavailable) binary coverage tracing.
Currently, UnTracer relies on a mix of black- and white-box binary instrumentation for constructing its two versions of the target binary.
For tracer binaries, we use Dyninst-based black-box binary rewriting~\cite{noauthor_dyninst_2018} to insert the forkserver and tracing infrastructure; for oracles, we re-purpose AFL's assembler front-end (\texttt{afl-as})~\cite{zalewski_american_2017} to insert the forkserver.
As discussed in Section~\ref{sec:implementation:forkserver}, our initial implementation used Dyninst to instrument the oracle binary, but we had to switch at \texttt{afl-as} due to unresolved performance issues.
Though instrumenting the oracle's forkserver at assembly-time requires assembly code access, we expect that inserting the forkserver is not a technical challenge for modern black-box binary rewriters~\cite{hawkins_zipr:_2017, wang_ramblr:_2017, wang_reassembleable_2015, bernat_anywhere_2011} or through function hooking (e.g., via \texttt{LD\_PRELOAD}~\cite{lopez_survey_2017}). 

\makeatletter{}\section{Related Work}

Two research areas orthogonal, but, closely related to coverage-guided tracing are improving test case generation, because improvements here increase the rate of coverage-increasing test cases and system optimizations, because they share the net outcome of improving overall fuzzer performance. 
We overview recent work in each area and relate those results back to coverage-guided tracing.

\subsection{Improving Test Case Generation}

Coverage-guided grey-box fuzzers like AFL~\cite{zalewski_american_2017} and libFuzzer~\cite{serebryany_continuous_2016} generally employ ``blind'' test case generation---relying on random mutation, prioritizing coverage-increasing test cases. 
A drawback of this strategy is stalled coverage, e.g., when mutation fails to produce test cases matching a target binary's \emph{magic bytes} (multi-byte strings or numbers) comparison operations. 
Research approaches this problem from several directions: Driller~\cite{stephens_driller:_2016} and QSYM~\cite{yun_qsym:_2018} use concolic execution (i.e., a mix of concrete and symbolic execution) to attempt to solve magic byte comparisons via symbolic path constraints. 
As is common with symbolic execution, exponential path growth becomes a limiting factor as target binary complexity increases. 
honggFuzz~\cite{swiecki_honggfuzz_2018} and VUzzer~\cite{rawat_vuzzer:_2017} both leverage static and dynamic analysis to identify locations and values of magic bytes in target binaries. 
Steelix~\cite{li_steelix:_2017} improves coverage by inferring magic bytes from lighter-weight static analysis and static instrumentation. 
Angora~\cite{chen_angora:_2018} incorporates byte-level taint tracking, outperforming Steelix's coverage on the synthetic LAVA datasets~\cite{dolan-gavitt_lava:_2016}.
However, despite seeing higher rates of coverage-increasing test cases, these fuzzers still face the overhead of tracing all generated test cases.

Instead of attempting to focus mutation on match magic byte comparisons all at once, an alternative set of approaches uses program transformation to make matching more tractable. 
AFL-lafIntel~\cite{noauthor_laf-intel:_2016} unrolls magic bytes into single comparisons at compile-time, but currently only supports white-box binaries. 
MutaGen~\cite{kargen_turning_2015} utilizes mutated ``input-producing'' code from the target binary for test case generation, but it relies on input-producing code availability, and faces slow execution speed due to dynamic instrumentation. 
T-Fuzz~\cite{peng_t-fuzz:_2018} attempts to strip target binaries of coverage-stalling code, but suffers ``transformational explosion'' on complex binaries.

Changes in test case mutation schemes have also offered potential workarounds to stalled coverage. FidgetyAFL~\cite{zalewski_afl-users_2016}, AFLFast~\cite{bohme_coverage-based_2016}, and VUzzer all prioritize mutating test cases exercising rare basic blocks. 
Ultimately, coverage-guided fuzzers identify coverage-increasing test cases by tracing the coverage of \emph{all} test cases. While such approaches decrease the number of test cases required to create a coverage-increasing test case, their rates of discarded test cases mean that coverage-guided tracing represents a performance improvement.

\subsection{System Scalability}

System scalability represents an additional focus of research on improving fuzzing.
AFL's execution monitoring component avoids overhead from repetitive \texttt{execve()} calls by instead using a fork-server execution model~\cite{zalewski_fuzzing_2014}. 
Xu et al.~\cite{xu_designing_2017} further improve AFL and libFuzzer's performance by developing several fuzzer-agnostic operating primitives. 
Distributed fuzzing has also gained popularity; Google's ClusterFuzz~\cite{google_clusterfuzz_2018} (the backbone of OSS-Fuzz~\cite{serebryany_oss-fuzz_2017}) allocates more resources to fuzzing by parallelizing across thousands of virtual machines. 
As these efforts aim to improve performance of \emph{all} fuzzers, they serve as complements to other fuzzing optimizations (e.g., coverage-guided tracing).

\makeatletter{}\section{Conclusion}

Coverage-guided tracing leverages the fact that coverage-increasing test cases are the overwhelmingly uncommon case in fuzzing by modifying target binaries so that they self-report when a test case produces new coverage. 
While our results show that the additional steps involved in coverage-guided tracing (namely, running the modified binary, tracing, and unmodifying based on new coverage) are twice as expensive as tracing alone, the ability to execute test cases at native speed, combined with the low rate of coverage-increasing test cases, yields overhead reductions of as much as 1300\% and 70\% for black- and white-box binaries, respectively. 
Applying coverage-guided tracing in hybrid fuzzing achieves 616\% and 79\% more test case executions than black- and white-box tracing-based hybrid fuzzing, respectively.
Thus, given that tracing consumes over 90\% of the total time spent fuzzing---even for fuzzers that focus on test case generation---reductions in tracing time carry over to fuzzing as a whole;

From a higher level, our results highlight the potential advantages of identifying and leveraging asymmetries inherent to fuzzing. 
Fuzzing relies on executing many test cases in the hopes of finding a small subset that are coverage-increasing or crash-producing. 
Even given recent attempts to reduce the number of discarded test cases, they are still the common case. 
Another opportunity is that most of the code itself is uninteresting, but must be executed to reach the interesting code. 
Thus, we envision a future where faster than full-speed execution is possible by finding ways to skip other ``uninteresting'' but common aspects of fuzzing.

\appendices

\section*{Acknowledgment}

We would like to thank our reviewers for helping us improve the paper.
We also thank Xiaozhu Meng from the Dyninst project and Insu Yun from the QSYM project for graciously assisting us in utilizing their software in our implementations.
Lastly, we thank Michal Zalewski for providing guidance on the inner workings of AFL.
This material is based upon work supported by the National Science Foundation under Grant No. 1650540.

\newpage{}

\bibliographystyle{IEEEtran}
\input{main.bbl}

\end{document}